\documentclass[lettersize,journal]{IEEEtran}
\pdfoutput=1
\usepackage{amsmath,amsfonts}
\usepackage{algorithmic}
\usepackage{algorithm}
\usepackage{array}
\usepackage[caption=false,font=normalsize,labelfont=sf,textfont=sf]{subfig}
\usepackage{textcomp}
\usepackage{stfloats}
\usepackage{url}
\usepackage{verbatim}
\usepackage{graphicx}
\usepackage{cite}
\usepackage{graphicx}
\usepackage{multirow}
\usepackage{booktabs} 
\usepackage{longtable}
\usepackage{array}
\usepackage{amsmath}
\usepackage{amssymb}
\usepackage{threeparttable}
\usepackage{hyperref}
\hypersetup{hidelinks,
	colorlinks=true,
	allcolors=black,
	pdfstartview=Fit,
	breaklinks=true}
\begin{document}

\title{Towards Robust Recommendation: A Review and an Adversarial Robustness Evaluation Library}

\author{Lei~Cheng,
        Xiaowen~Huang$^{\ast}$,
        Jitao~Sang,
        and~Jian~Yu,~\IEEEmembership{Member,~IEEE}
        
    \IEEEcompsocitemizethanks{
        \IEEEcompsocthanksitem Lei Cheng, Xiaowen Huang, Jitao Sang,  and Jian Yu are with the School of Computer Science and Technology, Beijing Jiaotong University, and also with the Beijing Key Lab of Traffic Data Analysis and Mining, Beijing Jiaotong University.
        (E-mail: \href{mailto:leicheng@bjtu.edu.cn}{leicheng@bjtu.edu.cn}; \href{mailto:xwhuang@bjtu.edu.cn}{xwhuang@bjtu.edu.cn}; \href{mailto:jtsang@bjtu.edu.cn}{jtsang@bjtu.edu.cn}; \href{mailto:jianyu@bjtu.edu.cn}{jianyu@bjtu.edu.cn})
        \IEEEcompsocthanksitem Lei Cheng, Xiaowen Huang and Jitao Sang are with the Key Laboratory of Big Data \& Artificial Intelligence in Transportation(Beijing Jiaotong University), Ministry of Education.
        }
         
    \thanks{$^{\ast}$corresponding author}

}

\markboth{Journal of \LaTeX\ Class Files,~Vol.~14, No.~8, August~2021}%
{Shell \MakeLowercase{\textit{et al.}}: A Sample Article Using IEEEtran.cls for IEEE Journals}


\maketitle

\begin{abstract}
Recently, recommender system has achieved significant success. However, due to the openness of recommender systems, they remain vulnerable to malicious attacks. Additionally, natural noise in training data and issues such as data sparsity can also degrade the performance of recommender systems. Therefore, enhancing the robustness of recommender systems has become an increasingly important research topic. In this survey, we provide a comprehensive overview of the robustness of recommender systems. Based on our investigation, we categorize the robustness of recommender systems into adversarial robustness and non-adversarial robustness. In the adversarial robustness, we introduce the fundamental principles and classical methods of recommender system adversarial attacks and defenses. In the non-adversarial robustness, we analyze non-adversarial robustness from the perspectives of data sparsity, natural noise, and data imbalance. Additionally, we summarize commonly used datasets and evaluation metrics for evaluating the robustness of recommender systems. Finally, we also discuss the current challenges in the field of recommender system robustness and potential future research directions.

\setlength{\parindent}{1em} 
Additionally, to facilitate fair and efficient evaluation of attack and defense methods in adversarial robustness, we propose an adversarial robustness evaluation library--ShillingREC, and we conduct evaluations of basic attack models and recommendation models. ShillingREC project is released at \url{https://github.com/chengleileilei/ShillingREC}.

\end{abstract}


\section{Introduction}
\label{sec:1}

\IEEEPARstart{W}{ith} the explosive growth of information and data on the Internet, recommender systems have become a crucial tool for addressing the problem of information overload. These systems excel at solving the problem of matching users with relevant items, making them widely applicable across various domains, including online shopping platforms, video content recommendations, social networks, news portals, and even in financial and healthcare systems.


While recommender systems offer a win-win situation for both users and businesses, it is evident that recommender systems are vulnerable to manipulation through malicious attacks. Kaya et al. (2023) \cite{kaya2023robustness} evaluated the robustness of three item-based multi-criteria recommender systems using the ``PIA'' attack model. The experimental results demonstrate that multicriteria top-n recommender systems are vulnerable to manipulation.  
In real-world scenarios, Amazon experienced a massive data breach involving fake reviews, which exposed more than 200,000 users and 13 million records, revealing widespread review manipulation on the platform\footnote{\url{https://www.techcentral.ie/data-breach-exposes-widespread-fake-reviews-on-amazon}}. Such attacks not only mislead consumers and damage platform credibility but also result in substantial economic losses. According to a report by CHEQ, fake reviews cause nearly \$152 billion in global economic damage annually\footnote{\url{https://www.saleswarp.com/fake-reviews-and-what-amazon-is-doing-about-it/}}.

Furthermore, some studies \cite{li2017collaborative,yang2022knowledge,chen2019improved} suggest that non-malicious factors such as sparse training data, sample imbalance, and unreasonable assumptions can also impact the performance of recommender systems. 
Therefore, the research on recommender systems has evolved beyond merely pursuing accuracy, with robustness becoming an important evaluation metric for recommender tasks. 

Therefore, in this comprehensive review, we have extensively collected relevant works on recommender system robustness, categorized and introduced them, including common attack methods, malicious attack detection methods, robust algorithms, and some non-adversarial robustness methods. Additionally, we have presented commonly used datasets and evaluation metrics in the task of recommender system robustness. Furthermore, we have implemented an adversarial robustness evaluation library, ShillingREC. ShillingREC integrates common attack models and recommendation models, enabling fair and efficient evaluation of these models. 

\subsection{Main Contributions}

The purpose of this review is to thoroughly review the literature related to recommender system robustness and further discuss the issues and directions in this field.  The main contributions of this review are summarized as follows: 
\begin{itemize}
\item We categorize the robustness of recommender systems from the perspective of whether they are vulnerable to adversarial attacks, distinguishing between adversarial robustness and non-adversarial robustness. We provide a comprehensive review of both classical and recent works on the robustness of recommender systems.
\item We compile commonly used datasets and evaluation metrics for the robustness of recommender systems and classify and introduce them accordingly.
\item We propose an adversarial robustness evaluation library, named ShillingREC, which integrates common attack models and recommendation models, enabling fair and efficient evaluation of these models. ShillingREC project is released at \url{https://github.com/chengleileilei/ShillingREC}.
\end{itemize}

\subsection{Difference with Existing Surveys}
Currently, there are several reviews \cite{deldjoo2021survey} that introduce the robustness of recommender systems from different perspectives. However, there is a lack of comprehensive reviews that systematically classify and organize the robustness of recommender systems. 




Deldjoo etal. (2021) \cite{deldjoo2021survey} discusses the latest advances in adversarial machine learning in recommender system security, primarily covering adversarial training techniques and GANS. However, this work only represents a branch of content related to recommender system robustness and does not provide a systematic investigation of recommender system robustness. Ji et al. (2020) \cite{JiShouling_2022} reviews the challenges currently faced by robustness in deep learning models and provides a systematic summary and scientific induction of existing research. However, this review is unrelated to the topic of recommender systems. Si and Li et al. (2020)\cite{si2020shilling} discusses shilling attacks in recommender systems and the corresponding defense mechanisms but does not conduct a systematic investigation of recommender system robustness from a higher perspective. Ge et al. (2022) \cite{ge2022survey} and Dai et al. (2022) \cite{fan2022comprehensive} provided comprehensive reviews of trustworthy recommender systems, covering aspects such as fairness, interpretability, robustness, and privacy protection. However, these works lack a detailed classification of robustness-related research in recommender systems and do not address non-adversarial robustness aspects.

In contrast, our work provides a comprehensive overview of the robustness of recommender systems from a broader perspective, encompassing both adversarial and non-adversarial robustness. We systematically review and introduce relevant literature from these two angles, while also revealing unresolved issues and future directions.

\begin{figure}[h]
  \centering
  \includegraphics[width=\linewidth]{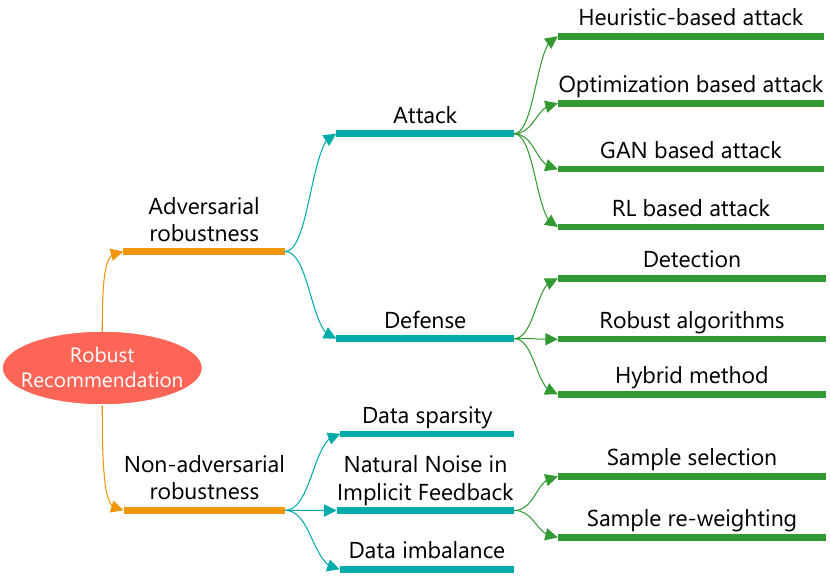}
  \caption{Recommender system robustness taxonomy diagram.}

  \label{fig:robust_taxonomy}
\end{figure}
\subsection{Structure of the Survey}
Section \ref{sec:2} of this review presents the taxonomy for the robust recommender systems. Section \ref{sec:3} focuses on the adversarial robustness of recommender systems, specifically addressing shilling attacks. In Section \ref{sec:3.1}, we introduce the concept and general definition of shilling attacks in the context of recommender systems. Section \ref{sec:3.2} categorizes shilling attack methods from different perspectives, followed by detailed discussions of specific methods in subsequent chapters. Section \ref{sec:4} delves into defense strategies against adversarial attacks on recommender systems, classifying and introducing various defense methods. Section \ref{sec:5} addresses the non-adversarial robustness of recommender systems. In Section \ref{sec:5.1}, we define non-adversarial robustness and discuss its underlying causes, followed by a presentation of methods to enhance the non-adversarial robustness of recommender systems in subsequent chapters. Section \ref{sec:6} introduces commonly used datasets for evaluating the robustness of recommender systems. Section \ref{sec:7} elaborates on the evaluation metrics commonly employed in assessing the robustness of recommender systems. Section \ref{sec:8} introduces the ShillingREC library for evaluating shilling attacks and defense mechanisms in recommender systems. Section \ref{sec:9} outlines the current challenges and future research directions in the robustness of recommender systems. Finally, a concluding section summarizes the key points discussed throughout the review.

\section{Taxonomy for Robust Recommendation}\label{sec:2}

Based on whether recommender systems are subjected to intentionally designed malicious attacks, we classify the robustness of recommender systems into adversarial robustness and non-adversarial robustness. The specific classification details are illustrated in Figure \ref{fig:robust_taxonomy}.

\subsection{Adversarial Robustness}
The emergence of robustness challenges in recommender systems stems from both model properties and external factors.

From a model perspective, CF uses group knowledge to work, which is central to its effectiveness, but it is also what makes CF so vulnerable. In CF, the actions of anyone can affect the recommendation results of others. And CF cannot detect malicious attackers naturally. Therefore, the attacker can easily influence the recommendation results to achieve the purpose of ``informal``. Hence, some studies \cite{kaya2023robustness} have confirmed that high-information attack models can be highly successful in user-based CF methods, while item-based methods tend to be more robust.

External factors analysis reveals that attackers may target recommender system configuration files with motives such as personal gain, commercial intrusion, system disruption, and orchestrating public opinion attacks on content recommendation platforms. Common tactics include manipulating ratings or comments within the recommender system to inflate or deflate the importance of specific items. For instance, elevating the ranking of one's own products or services while diminishing the reputation of competitors.

From a research perspective, the robustness of recommender systems against adversarial attacks can be divided into two parts: attack and defense. Looking from the defensive standpoint, the capability of recommender systems to withstand malicious attacks is termed as the robustness of recommender systems against adversarial attacks. Generally, the level of robustness can be measured by comparing the performance of recommender systems before and after malicious attacks. However, the research on algorithms for enhancing the robustness of recommender systems against adversarial attacks relies on interaction data with fake users, which is not readily available in real production environments. Therefore, to address the scarcity of data and improve the effectiveness of attacks, researching methods for generating adversarial attacks, also known as attack generation or ``adversarial attacks``, has become an important topic. Common approaches for generating adversarial attacks include heuristic methods, optimization-based methods, and model-based methods.

\subsection{Non-adversarial Robustness}
In addition to malicious attacks targeting recommender systems, several non-malicious factors can also affect their performance, such as data sparsity, natural noise in training data, and data imbalance. The ability of recommender systems to withstand such negative effects is referred to as their non-adversarial robustness. A detailed discussion of the non-adversarial robustness of recommender systems is provided in Section \ref{sec:5}.

\begin{figure}[h]
\setlength{\abovecaptionskip}{-0.1cm}
  \centering
  \includegraphics[width=0.9\linewidth]{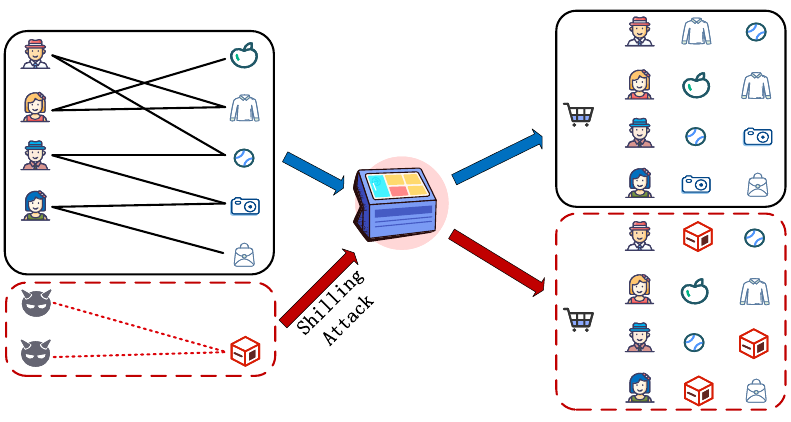}
  \caption{Diagram of Shilling attack. Blue arrows indicate the recommendation process unaffected by adversarial attacks, while red arrows represent the recommendation process in the presence of adversarial attacks.}
  \label{fig:shilling_attack}
  \vspace{-1 em}
\end{figure}

\section{Attack in Adversarial Robustness}\label{sec:3}
\subsection{Concept and General Definition}\label{sec:3.1}

The CF recommender system relies on ``collective intelligence'', where item calculation is based on information from other items. However, this characteristic makes CF recommender systems vulnerable to distinguishing genuine users from fake ones. Moreover, these systems significantly impact the revenue of businesses or platforms. Consequently, malicious entities, whether companies or users, can manipulate CF methods to their advantage by injecting malicious data. They may select target products and manipulate data by adding fake users and interaction records to alter the popularity of these items, known as shilling or profile injection attacks, as illustrated in Figure \ref{fig:shilling_attack}.

\begin{figure}[h]
  \centering
  \includegraphics[width=0.4\linewidth]{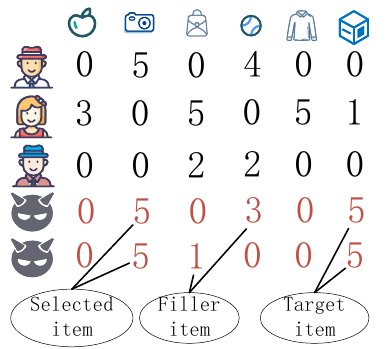}
  \caption{
The foundational parameters for shilling attack.}
  \label{fig:shilling_data}
    \vspace{-1 em}
\end{figure}

From the perspective of the specific implementation of shilling attacks, taking targeted shilling attacks as an example, there are three basic parameters: attack size, target item, and filler item. The attack size refers to the number of fake users injected into the recommender system, the target item is the specific object of the shilling attack, which can be one or more items, and the filler item is other interaction items of fake users. The purpose is to conceal the attack behavior and enhance the stealthiness of the attack. The filler item size is usually set to the average number of items in the dataset. In addition, some attack models manually select certain items as ``selected items'' and assign them specific scores to amplify the destructiveness and concealment of the attack. Taking rating data as an example, the basic parameters of shilling attacks are depicted in Figure \ref{fig:shilling_data}.

\subsection{Taxonomy}\label{sec:3.2}

O'Mahony et al. (2002) \cite{OMahonyb} initially discussed attacks on CF. Subsequently, O'Mahony et al. (2004) \cite{OMahonya} formally defined several foundational attack strategies. However, the landscape of attack and defense techniques has been an ongoing, reciprocal process in recommender systems. Following the introduction of shilling attacks in recommender systems, corresponding defense methods were proposed. Subsequently, more advanced attack techniques have been studied, leading to a series of methods for both attacking and defending recommender systems.

In this section, we have compiled and organized common and representative adversarial attacks at various stages, categorizing them according to different criteria.

\subsubsection{Attack Target}
Firstly, according to the objective of adversarial attacks, they can be classified into targeted attacks and non-targeted attacks. Specifically, targeted attacks can be further divided into two categories: push and nuke\cite{kaya2023robustness}. Push attacks involve injecting fake configuration files to increase the popularity of a specific item, with the aim of improving the ratings or exposure of the target item. In contrast, nuke attacks involve injecting fake configuration files to decrease the popularity of a specific item. On the other hand, non-targeted attacks do not focus on specific items but rather aim at the entire recommender system. The goal of nontargeted attacks is to disrupt the overall performance of the target recommender system, thereby reducing the overall satisfaction of platform users. In terms of frequency of use and practical scenarios, targeted attacks generally have a relatively lower implementation difficulty and are more applicable to real-world scenarios, making them more prevalent. In contrast, non-targeted attacks, due to their high requirements on attackers, are less commonly applied in practical scenarios. However, they are often used to evaluate the robustness of a particular recommendation model.

\subsubsection{Attack Knowledge}
The generation of deceptive data by attackers typically requires some prior knowledge, such as knowledge of the victim's model, model parameters, and training data. Based on the knowledge required for poisoning attacks, they can be classified as white-box attacks, gray-box attacks, and black-box attacks. White-box attacks involve the attacker knowing the structure and parameters of the victim's model. For example, Li et al. (2016)\cite{liDataPoisoningAttacks2016} proposed the PGA and SGLD poisoning attack algorithms for matrix factorization (MF), and Fang et al. (2018) introduced the SRWA poisoning attack algorithm for graph recommender systems \cite{fangPoisoningAttacksGraphBased2018}.Gray-box attacks refer to cases where the victim's model's recommendation algorithm is known, but its parameters are unknown. For instance, Wu et al. (2021) presented the TrialAttack, which, with only partial knowledge, obtains some data through means like web scraping to train a local model simulator for conducting poisoning attacks \cite{wuTripleAdversarialLearning2021}.Black-box attacks involve attackers lacking knowledge about both the victim model's structure and its parameters. For example, Zhang et al. (2020) proposed LOKI, and Song et al. (2020) introduced PoisonRec, using reinforcement learning techniques to achieve poisoning attacks on black-box recommender systems \cite{zhangPracticalDataPoisoning2020,songPoisonRecAdaptiveData2020}.

\begin{figure}[h]
  \centering
  \includegraphics[width=\linewidth]{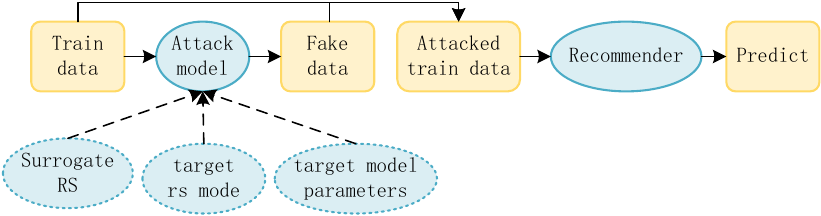}
  \caption{Recommender system shilling attack workflow diagram.}
  \label{fig:shilling_flow}
    \vspace{-1 em}
\end{figure}

Although the required knowledge may vary among different shilling attacks, the workflow for shilling attacks is generally similar. Shilling attack models typically involve acquiring training or interaction data, generating fake interaction data, merging the fake data with the original training data, and injecting them into the target recommender system to complete the attack, as illustrated in Figure \ref{fig:shilling_flow}.

\subsubsection{Attack Method}
As recommender systems and machine learning technologies continue to advance, various adversarial attack methods have emerged. Based on the attack approaches, adversarial attacks can be categorized into heuristic-based methods, optimization-based methods, GAN-based methods, and reinforcement learning-based methods. Although the emphasis may vary among these methods, the overall goal remains the pursuit of the invisibility and destructiveness of adversarial attacks. In the next section, we will use targeted push attacks as an example to introduce specific classification rules and related works. 

We have summarized the investigation work based on different implementation methods of shilling attacks and the required knowledge for executing shilling attacks. Please refer to Table \ref{tab:attackmethod} for the details.

\begin{table*}
\setlength{\abovecaptionskip}{-0.05cm} 
\setlength{\belowcaptionskip}{-0.2cm} 
    \centering
    \caption{Statistics of Shilling Attack Approaches}

    \resizebox{\linewidth }{!}{
\begin{tabular}{ccccccccccc}
    \toprule
    \multicolumn{2}{c}{\multirow{2}{*}{Category}} & \multirow{2}{*}{Method} & \multirow{2}{*}{Data type} & \multirow{2}{*}{RS task} & \multirow{2}{*}{Rank strategy} & \multirow{2}{*}{Metric} & \multirow{2}{*}{\begin{tabular}[c]{@{}c@{}}Surrogate\\ model\end{tabular}} & \multicolumn{2}{c}{Attack knowledge} & \multirow{2}{*}{Time} \\ \cline{9-10}
    \multicolumn{2}{c}{} &  &  &  &  &  &  & \begin{tabular}[c]{@{}c@{}}Target\\ recommend\\ model\end{tabular} & \begin{tabular}[c]{@{}c@{}}Target\\ model\\ parameters\end{tabular} &  \\ \midrule
    \multirow{5}{*}{Heuristic} &  & Random Attack & explicit & rating & -- & -- &  &  &  & 2004 \\
     &  & Average attack & explicit & rating & -- & -- &  &  &  & 2004 \\
     &  & Bandwagon attack & explicit & rating & -- & -- &  &  &  & 2007 \\
     &  & Love/hate attack & explicit & rating & -- & -- &  &  &  & 2007 \\
     &  & Segment Attack & explicit & rating & -- & -- &  &  &  & 2005 \\ \midrule
    \multirow{8}{*}{Optimization} & \multirow{5}{*}{Single level} & Co-visitation & implicit & rank & -- & AST &  &  &  & 2017 \\
     &  & PGA,SGLD & explicit & rating & -- & RMSE Rating &  & \checkmark & \checkmark & 2016 \\
     &  & SRWA & explicit & rank & pointwise & HR &  & \checkmark & \checkmark & 2018 \\
     &  & TNA & explicit & rank & pointwise & HR &  & \checkmark & \checkmark & 2020 \\
     &  & DL Attack & explicit & rank & pairwise & HR &  & \checkmark & \checkmark & 2020 \\ \cline{2-11} 
     & \multirow{3}{*}{Bi-level} & AIA/RevAdv & implicit & rank & pointwise & HR & \checkmark &  &  & 2020 \\
     &  & RAPU & implicit & rank & pairwise & HR & \checkmark &  &  & 2021 \\
     &  & Infmix & explicit & rank & pointwise & HR &  &  &  & 2023 \\ \cline{2-11} 
     & Cross domain & PC-Attack & implicit & rank & pairwise & HR NDCG &  &  &  & 2023 \\ \midrule
    \multirow{6}{*}{GAN} &  & DCGAN & explicit & rank & pointwise & HR &  & \checkmark & \checkmark & 2019 \\
     &  & AUSH & explicit & rank & pointwise & HR &  &  &  & 2020 \\
     &  & RecUP & explicit & rank & pointwise & HR &  &  &  & 2021 \\
     &  & GOAT & explicit & rank & pairwise & HR Precision NDCG &  &  &  & 2021 \\
     &  & TrialAttack & explicit & rank & pointwise & HR NDCG & \checkmark & \checkmark &  & 2021 \\
     &  & Leg-UP & explicit & rank & pointwise & HR & \checkmark &  &  & 2022 \\ \midrule
    \multirow{3}{*}{RL} &  & LOKI & implicit & rank & pairwise & HR & \checkmark &  &  & 2020 \\
     &  & PoisonRec & implicit & rank & pointwise & RecNum &  &  &  & 2020 \\
     &  & CopyAttack & implicit & rank & pairwise & HR NDCG &  &  &  & 2021 \\
     \bottomrule
    \end{tabular}
     }
    \label{tab:attackmethod}
\end{table*}

\subsection{Heuristic-based Attack}\label{sec:3.3}
Heuristic-based attacks generate fake profiles using predefined fixed strategies and statistical features. These methods are straightforward, resource-efficient, and broadly applicable, allowing them to target various recommendation models and contexts. Common heuristic-based attack techniques include random attacks, average attacks, bandwagon attacks, love/hate attacks, and segment attacks.

In the random attack strategy \cite{lamShillingRECommenderSystems2004}, the first step is to assign the highest (push) or lowest (nuke) rating to the target item for all fake users. Then, the global rating mean and standard deviation are computed to establish a normal distribution for the rating scores. Finally, each fake user randomly selects n filler items, and the ratings for these items are randomly sampled from the constructed normal distribution.

In the average attack \cite{lamShillingRECommenderSystems2004}, the first step is to assign the highest (push) or lowest (nuke) rating to the target item for all fake users. Next, the mean and standard deviation of the ratings are calculated for each item to establish a normal distribution for its rating scores. Finally, each fake user randomly selects n filler items, and ratings are assigned by sampling from the normal distribution of the corresponding item's rating scores.

Bandwagon Attack \cite{mobasherTrustworthyRecommenderSystems2007}, or Popular Attack, resembles random attacks. It assigns maximum or minimum ratings to target items for all fake users and calculates the global mean and standard deviation to create a normal distribution of rating scores. The key distinction is in filler item selection: the Bandwagon Attack mixes a proportion of popular items (with higher average ratings) with random items, sampling rating scores from the established normal distribution.

The Love/Hate Attack \cite{mobasherTrustworthyRecommenderSystems2007} operates without prior knowledge of the data. In a push attack, all fake users are assigned the maximum rating for the target item, while each fake user randomly selects n filler items and assigns them the minimum rating. Although Love/Hate Attack is effective against user-based collaborative filtering, its straightforward filler strategy makes it relatively easy to detect.

Segment Attack \cite{burkeSegmentbasedInjectionAttacks2005} is a covert attack method designed to effectively recommend a target item to interested users. It assigns the highest rating to the target item for all fake users and selects several similar items as the selected items, also giving them the highest ratings. In addition, n filler items are randomly chosen and assigned the lowest ratings. Although Segment Attack requires more prior knowledge than other methods, its rating patterns closely resemble those of legitimate users, enhancing its subtlety.

While heuristic methods are simple, fast, and widely applicable, their implementation principles suggest that they may lack precision, leading to suboptimal attack effectiveness. Moreover, the use of a singular generation strategy in heuristic attacks can make them prone to detection, reducing their stealthiness.

\subsection{Optimization-based Attack}\label{sec:3.4}
Optimization-based methods do not employ fixed strategies to generate fake data; instead, they investigate modeling adversarial attacks as optimization tasks and subsequently apply optimization strategies for resolution. Based on the configuration of the optimization task, optimization-based methods can be categorized into Single-level optimization methods and Bi-level optimization methods.

\subsubsection{Single-level Optimization}
The Single-level optimization method defines targeted attacks as a single-layer optimization problem, with the objective of maximizing the attack loss, i.e.:
\begin{equation}\min_{\widehat{X}}\mathcal{L}_{\mathrm{adv}}(\mathbf{R}_{\theta}),\end{equation}
Therefore, to address this optimization problem, the Single-level optimization method generally requires an understanding of the structure and parameters of the victim model.


The pioneering work in optimization-based attacks was introduced by Li et al.\cite{liDataPoisoningAttacks2016}, presenting the PGA and SGLD methods. Specifically, PGA focuses on MF models, defining different loss functions based on various attack objectives. It then optimizes these loss functions according to the model's prediction results. Additionally, Fang et al. (2018) \cite{fangPoisoningAttacksGraphBased2018} proposed the SRWA method for targeted attacks on graph recommender systems. SRWA employs random walks for probability prediction, approximating the hit rate with the steady-state probabilities of items. The optimization strategy maximizes the hit rate of the attack target.

To simplify the optimization algorithm, Fang et al. (2020) \cite{fangInfluenceFunctionBased2020} introduced TNA. TNA utilizes influence functions to select a subset S of users who have an impact on the target item. It considers the items corresponding to users in S as templates, optimizes the scores of the user's interaction items based on the attack objective, selects high-scoring items as filler items according to the attack budget, and sets filler item scores based on the average attack strategy. Finally, it determines the specific score for the target item based on the attack objective.

Inspired by SRWA, Huang et al. (2021) \cite{Huang2021} proposed a poisoning attack method for deep learning-based recommender systems called DL Attack. DL Attack utilizes a loss function to approximate the hit ratio and constructs a poison model to simulate a compromised deep learning recommender system. Ultimately, it builds filler items and their ratings for fake users through a two-stage training process of the poison model, employing a one-by-one approach.

\subsubsection{Bi-level Optimization}
The Single-level optimization method assumes that the model parameters of the recommender system are fixed and do not change during the optimization process. This unrealistic assumption overlooks the fact that model parameters may change with the poisoned data during the optimization process. Such an unreasonable assumption can lead to significant biases in threat estimation \cite{wuInfluenceDrivenDataPoisoning2023}.To address this issue, Bi-level optimization methods have been proposed. These methods typically do not require knowledge of the specific parameters of the target recommender system. Instead, they construct a local proxy recommendation model. The optimization objective is usually defined as follows:
\begin{equation}
\begin{split}
& \min_{\widehat{X}}\mathcal{L}_{\mathrm{adv}}(R_{\theta^*}), \\
& \text{s. t.}\quad\theta^*=\arg\min_{\theta}\left(\mathcal{L}_{\mathrm{train}}(X,R_\theta)+\mathcal{L}_{\mathrm{train}}(\widehat{X},\widehat{R}_\theta)\right),
\end{split}
\end{equation}
where $ X $ denotes the original data of the recommendation model, $ \widehat{X} $ represents the injected fake data into the recommender system, $ \theta $ denotes the parameters of the surrogate model, and $ R_\theta $ and $ \widehat{R}_\theta $ denote the outcomes obtained by training the recommendation model with parameters $ \theta $ on the original and fake data, respectively.

During the optimization process, both the attack model and the proxy model undergo parameter updates, ultimately achieving precise targeted attacks. This bi-level optimization approach ensures that the models evolve together, capturing the dynamics of the recommender system and allowing for accurate adversarial attacks.


The classic model of the Bi-level optimization method is AIA proposed by Tang et al. (2020) \cite{tangRevisitingAdversariallyLearned2020}. In AIA, the authors review the bi-level optimization problem of the surrogate model and propose a time-efficient and resource-efficient solution based on bi-level optimization strategies. Subsequently, Zhang et al. (2021) \cite{zhangDataPoisoningAttack2021} proposed a bi-level optimization framework RAPU, which can achieve cross-level adversarial attacks under the constraint of limited training data. This framework utilizes a probabilistic generative model to identify users and items with sufficient and minimally disturbed interaction counts and generates fake interaction data based on this foundation. The data generated by this method are implicit feedback data.Bi-level optimization methods typically require simultaneous optimization of both the attack model and the surrogate recommendation model, resulting in substantial computational costs. Consequently, Wu et al. (2023) \cite{wuInfluenceDrivenDataPoisoning2023} introduced Infmix as a solution. Infmix employs an influence-based threat estimator module to assess the harm of fake users. This module evaluates the harm of fake users without the need to retrain the recommender system. Combined with the distribution-agnostic generator Usermix, Infmix is capable of generating concealed and adversarial fake data.

Hence, when it comes to attack methods, optimization-based approaches offer greater accuracy and the ability to be customized compared to heuristic methods. These approaches can effectively focus on specific target models, enabling the use of different optimization strategies based on specific situations, thus increasing the flexibility of the attacks. However, the use of optimization-based attack methods also comes with higher computational complexity, especially in scenarios involving large-scale datasets and complex models. Furthermore, the effectiveness of these methods may be weakened by defense mechanisms such as adversarial training, which can reduce the efficiency of attacks.

\subsection{GAN-based Attack}\label{sec:3.5}


GAN (Generative Adversarial Network) is a generative model composed of a generator and a discriminator. These two networks are trained through adversarial learning, forming a zero-sum game. The generator takes random noise as input, and its learning objective is to generate data from the latent space to approximate the distribution of real data. On the other hand, the discriminator is responsible for distinguishing between the data generated by the generator and real data. By alternately optimizing these two networks, GAN achieves an optimal balance between the generator and the discriminator, terminating at a saddle point that represents the minimum value for the generator and the maximum value for the discriminator.

The mathematical formulation of GAN is as follows.

\begin{equation}
\begin{aligned}
& \min_G\max_DV(D,G) = \\ 
& \mathbb{E}_{x\sim p_\text{data}(x)}[\log D(x)]+\mathbb{E}_{z\sim p_z(z)}[\log(1-D(G(z)))],
\end{aligned}
\end{equation}

where $ p_{\text{data}}(x) $ denotes the distribution of real data, $ p_z(z) $ denotes the distribution of the latent space input of the generator, $ D(x) $ denotes the output of the discriminator for real data and $ G(z) $ denotes the generated result of the generator for the latent variable $ z $. This adversarial learning approach endows GAN with powerful capabilities.

In the domain of adversarial attacks on recommender systems, optimization-based adversarial attack methods often focus excessively on the harmfulness of the attack, neglecting the importance of concealing the attack. To achieve a covert and widely applicable adversarial attack, some studies have introduced GAN technology \cite{goodfellow2014generative} into the field of adversarial attacks. The DCGAN\cite{christakopoulouAdversarialAttacksOblivious2019} method, proposed by Christakopoulou and Banerjee, is the first to utilize GAN technology for adversarial attacks. Specifically, the DCGAN generator takes the random noise input into convolutional units and produces fake profiles that conform to the distribution of real users. The discriminator is employed to identify the authenticity of the generated data. Additionally, to address the challenge of directly obtaining gradients for optimising attack objectives, zero-order optimization methods are used to estimate the gradients of the attack target, ultimately achieving a highly effective covert adversarial attack.

Subsequently, to reduce the training cost associated with GANs generating fake profiles, Lin et al. (2020) introduced AUSH\cite{linAttackingRecommenderSystems2020}. AUSH utilises sampled template users to construct ratings for filler items, resulting in lower training costs compared to the random initialisation strategy in DCGAN. To further enhance the concealment and harmfulness of the attack, Zhang et al. (2021) proposed a GAN-based adversarial attack framework called RecUp\cite{zhangAttackingRecommenderSystems2021}. RecUp designs an HRGAN module with a threat loss function, enabling the generator to produce both reasonable and harmful users by guiding the threat function. Similarly, to address the imbalance between aggressiveness and concealment, Wu et al. (2021) introduced an end-to-end adversarial attack framework named TrialAttack \cite{wuTripleAdversarialLearning2021}. Given input noise, TrialAttack can directly generate malicious users through triple adversarial learning involving a generator, discriminator, and influence module.

Additionally, Wu et al. (2021) presented an adversarial attack method called GOAT\cite{wuReadyEmergingThreats2021}, specifically designed for graph recommender systems. GOAT utilises i-i graph sampling to assist GAN in training to generate as realistic fake users as possible, and sets specific ratings for the target item based on the attack target. Lin et al. (2022) implemented an improved version of AUSH called Leg-UP\cite{linShillingBlackBoxRecommender2022}. Leg-UP uses a proxy recommendation model to assist with the training of the generator in GAN. Specifically, the score for the target item is optimized with the assistance of the proxy model based on the attack loss function, improving the transferability and concealment of the attack.

In summary, the advantages of adversarial attack methods based on GAN lie in the generation of more realistic users compared to heuristic and optimization-based methods. Additionally, some approaches enable end-to-end training, enhancing the generation of users with more significant attack impact. However, GAN models are prone to overfitting and carry the risk of model collapse during the training process. Moreover, when dealing with large-scale datasets, the training of GAN models may demand considerable computational resources and time.

\subsection{RL-based Attack}

To simulate more realistic attack scenarios, some studies employ reinforcement learning techniques for adversarial attacks on recommender systems. LOKI, proposed by Zhang et al. (2020) \cite{zhangPracticalDataPoisoning2020}, utilizes a proxy recommendation model and reinforcement learning techniques to achieve poisoning attacks on black-box recommender systems. The learning process of the model involves estimating the influence function to inject samples and observe their impact on the recommendation results, eliminating the need for retraining the proxy model.Another framework, PoisonRec, introduced by Song et al. (2020) \cite{songPoisonRecAdaptiveData2020}, presents a reinforcement learning-based adversarial attack. PoisonRec models the attack behavior as a Markov decision process, where the attack agent injects fabricated data into the recommender system. In a strict black-box setting, PoisonRec optimizes its attack strategy through reward signals.In the same year, Fan et al. (2021) \cite{fanAttackingBlackboxRecommendations2021} proposed the CopyAttack framework, also based on reinforcement learning. CopyAttack employs a policy gradient network to select user profiles from the source domain and refines them using a masking mechanism before injecting them into the target domain. After injecting fake data, attackers utilize spy users to evaluate the effectiveness of the injected fake data and define corresponding rewards.

In summary, reinforcement learning-based adversarial attack methods do not require knowledge of the actual model's structure and parameters, making them well-suited for black-box attack scenarios and adaptable to different recommender systems. They also possess a certain level of interpretability. However, reinforcement learning models still face challenges such as low data efficiency and learning instability.

\section{Defense in Adversarial Robustness}\label{sec:4}
With the advancement of recommender system attack methods, corresponding defense techniques have also emerged, becoming an indispensable part of recommender systems. We have transitioned from an era solely focused on optimising recommendation accuracy to a more diverse landscape, emphasising robustness, diversity, fairness, and various other dimensions. The robustness of recommendation now plays a significant and pivotal role within this evolving landscape.


\begin{figure}[h]
\setlength{\abovecaptionskip}{-0.1cm}
  \centering
  \includegraphics[width=\linewidth]{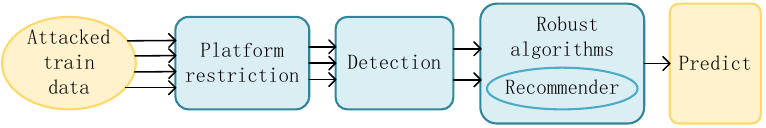}
  \caption{Diagram of defense in adversarial robustness.}
  \label{fig:defense}
    \vspace{-0.8 em}
\end{figure}

To counter the threat of shilling attacks and establish defense mechanisms within the entire workflow of recommender systems, three key approaches can be considered:

Firstly, it is possible to implement a rating threshold to prevent malicious reviews by bots. This could involve the introduction of validation mechanisms for rating operations on the recommendation platform. For example, users might need to pass certain verification steps before their ratings are considered valid.

Secondly, during the training phase of the recommender system, it is essential to analyze shilling attack characteristics and construct malicious attack detectors. These detectors can be used to filter out clean training data, ensuring that the recommendation model is trained on legitimate and untainted data, thus reducing susceptibility to manipulative attacks.

Lastly, leveraging various robust algorithms, it is possible to develop robust recommendation models. These models are designed to minimize noise and learn genuine user representations during the learning process. By incorporating robustness measures into the model's architecture, it becomes more resilient to shilling attacks and other adversarial influences.The schematic diagram illustrating the robustness defense mechanism is depicted in Figure \ref{fig:defense}.

This section provides a comprehensive overview of both detection-based methods and robustness algorithm-based approaches. As for the method involving the establishment of rating thresholds, which falls under system development, it will not be discussed in this paper.
\subsection{Detection}
According to the research findings, detection-based methods can be categorized based on the need for supervised data into three categories: supervised methods, unsupervised methods, and semi-supervised methods.
\subsubsection{Supervised Methods}
Supervised detection methods involve training a classifier based on labeled data to assess the likelihood that a user is a fake user, using features such as user ratings. Common approaches include SVM-based classifiers and CNN-based classifiers.

For instance, Zhou et al.\cite{zhou2016svm} proposed an SVM-based shilling attack detection method called SVM-TIA. This method trains a supervised SVM classifier and employs the Borderline-SMOTE method to balance the dataset while using target-item analysis to reduce false positive rates. Tong et al.\cite{tong2018shilling} introduced a CNN-based shilling attack detector for social-aware network collaborative recommender systems, known as CNN-SAD. CNN-SAD consists of a convolutional layer and a pooling layer, using sigmoid as the activation function. This represents the first application of deep learning techniques to detect recommendation attacks. Ebrahimian et al.\cite{ebrahimian2020detecting} introduced a mixed model that uses both CNN and RNN to detect shilling attacks, including CNN-LSTM and CNN-GRU. These methods can directly extract features from rating data and model the temporal and spatial information in the recommender system ratings, achieving end-to-end detection capabilities.

\subsubsection{Unsupervised Methods}
In the field of recommender systems, the majority of datasets do not come with labels indicating whether the data is part of a malicious attack. In supervised detection tasks, a common assumption is that the interactions in the dataset are genuine, and researchers manually inject malicious data into the training dataset to train the model. However, manually crafted malicious data may not accurately simulate real-world attacks in recommender systems. Additionally, without labels, it's challenging to identify whether unlabeled data corresponds to malicious behavior. As a result, some research has turned to unsupervised techniques like Principal Component Analysis (PCA) and clustering for shilling attack detection.

For example, Mehta and Nejdl (2009)\cite{mehta2009unsupervised} proposed a PCA-based detection method called PCA-VarSelect. PCA-VarSelect detects attack profiles by calculating covariance between users. This method is an early representative of unsupervised techniques and exhibits good detection performance. Zhang et al. (2015)\cite{zhang2015catch} introduced a method called CBS, which leverages the spread of fraudulent behavior to distinguish attackers. This approach does not require knowledge of the attack strategy but does need some labeled users as seeds to initiate the detection process. Hao and Zhang (2021)\cite{hao2021unsupervised} presented an unsupervised shilling attack detection method called DC-EDM, which combines deep learning and community detection. DC-EDM constructs a user graph based on the similarity of user behaviors, employs stacked denoising autoencoders to extract robust features, generates multiple clustering results, and then reconstructs the user graph. Finally, it uses community discovery methods and associated metrics to detect shilling attacks.

These unsupervised techniques are valuable in scenarios where labeled data for attack detection is scarce or unavailable, providing alternative approaches to identify malicious behavior in recommender systems.
\subsubsection{Semi-supervised Methods}
Semi-supervised methods combine the strengths of both supervised and unsupervised learning approaches, allowing for the utilization of a small amount of labeled data to identify malicious attack data.

For instance, Wu et al. (2012) introduced the semi-supervised learning system HySAD\cite{wu2012hysad}. HySAD leverages both labeled and unlabeled user profiles for multi-class modeling. It constructs traditional feature-trained Naive Bayes classifiers for labeled users and then employs the EM-$\lambda$ algorithm to detect unlabeled users using these classifiers. Zhou and Duan (2021) proposed a semi-supervised detection method called SSADR-CoF\cite{zhou2021semi}. SSADR-CoF first extracts a series of user rating behavior pattern features using window partitioning and rating behavior statistics. It then initializes the classifier using a small number of labeled user profiles and assigns labels to unlabeled user profiles using this classifier. Hao et al. (2023) addressed model-generated shilling and group attacks with a two-stage approach called DHAGCN\cite{hao2023detection} for attack detection. Firstly, they performed clustering based on user features and partially labeled user nodes based on the distribution around cluster centroids. Then, they constructed a GCN-based detector that analyzed the labeled users and their neighbors using spectral convolution for advanced representation.

\subsection{ Robust Algorithms}
While detection-based methods can help identify attack features and purify data before training recommendation models, they do have certain limitations. Firstly, detection-based methods cannot completely filter out malicious attacks, and secondly, in pursuit of higher recall rates, they may unavoidably remove genuine data, leading to issues such as data sparsity.

In light of these limitations, robust algorithms for recommender systems have garnered significant research attention. Robust algorithms aim to learn user representations that capture genuine preferences even in the presence of noisy data. Based on my research, this section categorizes mainstream recommender system robustness methods into three classes: model-based methods, adversarial training-based methods, and trust-aware recommendation methods.

\subsubsection{Model Based}
The model-based approach to enhancing robustness refers to methods aimed at improving the robustness of recommender systems by optimizing or refining the model's structure and solving techniques.

Due to MF techniques being one of the most successful and widely applied technologies in recommender systems, a multitude of robust algorithms have emerged for MF.

For instance, Cheng and Hurley (2010)introduced LTSMF\cite{cheng2010robust} , which utilizes least squares for MF. Experimental results indicate that LTSMF exhibits stronger robustness compared to MF methods based on M-estimation. Yi et al. (2014) proposed the RCR-KTM method\cite{yi2014robust}, which calculates user suspicion through neighborhood recommendations and combines it with MF based on Tukey's M-estimator for recommendation. Yu et al. (2017) presented a robust recommendation method called RRA-KMF\cite{yu2017novel}, which is based on kernel MF. RRA-KMF constructs a robust kernel MF model without square error terms using the rating matrix and kernel distances. Fadaee et al. (2018) introduced Chiron\cite{fadaee2018chiron}, a recommendation method that combines neighborhood-based and MF approaches. Chiron leverages both neighborhood information and MF for robust recommendations. Wu et al. (2020) proposed SL1-LF\cite{wu2020robust}, which replaces the L2 norm in MF with a smooth L1 norm latent factor model. Experimental results demonstrate that the SL1-LF model exhibits robustness against anomalous data. These robust algorithms tailored to MF contribute significantly to enhancing the robustness of recommender systems.


Some researchs \cite{Chen2021,Chen2020,Zhu2019} suggest that GNN-based recommender systems are vulnerable to malicious noise due to their recursive message passing schema. To address this issue, certain works have introduced strategies such as graph regularization and contrastive learning to enhance the robustness of GNN-based recommendation models.

For example, Chen et al. (2021) proposed the Structured Graph Convolutional Network (SGCN) \cite{Chen2021}, which enhances robustness against incomplete and noisy user-item graphs. SGCN achieves this by introducing random binary masks at each layer to prune noisy edges and applying nuclear norm regularization to maintain the low-rank characteristics of the graph.
Wu et al. (2021) proposed Self-Enhanced Graph Learning (SGL)\cite{wu2021self}, which supplements classical recommendation supervision tasks with auxiliary self-supervised tasks to strengthen node representation learning. Specifically, SGL generates multiple views of a node and maximizes consistency between different views of the same node through self-supervised learning. 
Yang et al. (2022)\cite{yang2022knowledge} leveraged additional signals from knowledge graphs to supervise cross-graph contrastive learning. This approach improved bias reduction for user-item pairs during gradient descent and mitigated noise. It demonstrated good performance in scenarios involving sparse interactions, long-tail data, and noise. 
Zhu et al. (2023)\cite{zhu2023knowledge} introduced the Knowledge-Refined Denoising Network (KRDN), which combines adaptive knowledge refinement and contrastive denoising while removing irrelevant relationships from knowledge graphs and noise interactions from implicit feedback. In the knowledge refinement part, an adaptive pruning strategy is employed to extract high-quality triplets for additional knowledge in recommendations. In the contrastive denoising part, noise is identified based on the difference between collaborative signals and knowledge signals to learn users' true preferences. 
Fan et al. (2023)\cite{fan2023mutual} improved contrastive learning losses by applying Wasserstein distance measurements to quantify mutual information between enhanced sequences, thus enhancing the robustness of sequence recommendations.

In addition to the aforementioned approaches, denoising autoencoders have opened a new avenue for robust recommender algorithms. Wu et al. (2016) introduced Collaborative Denoising Autoencoders (CDAE) [52] to mitigate random noise in observed interactions. They employed several linear layers to reconstruct the original data, enhancing the model’s resistance to noise.

\subsubsection{Adversarial Training}



Adversarial training in recommender systems involves adding adversarial perturbations to the model to enable it to learn more robust representations. Specifically, adversarial training is a minimax game in which the training process maximizes the loss induced by adversarial perturbations while minimizing the loss associated with the recommendation task. This process helps the model learn to capture genuine user preferences even in the presence of noise, thereby enhancing recommendation robustness.

The pioneering work in adversarial training for recommender systems is the APR\cite{he2018adversarial} method proposed by He et al. (2018), which implements adversarial training based on Bayesian Personalized Ranking (BPR) loss for MF. Subsequently, Yuan et al. (2019) introduced a general adversarial training framework called ACAE\cite{yuan2019adversarial} for neural network-based recommender system models. Chen and Li (2019) presented an adversarial training method, ATF\cite{chen2019adversarial}, based on context-aware tensor factorization.

The aforementioned methods achieve adversarial training by introducing small perturbations to parameters. However, this approach does not align with the actual poisoning mechanism in recommender systems. Therefore, Wu et al. (2021) introduced Adversarial Poisoning Training (APT)\cite{wu2021fight}, which simulates the poisoning process by injecting fake users (ERM users). These fake users aim to minimize empirical risk to construct robust recommendations.

In addition, some adversarial training-based methods have been proposed for multimedia recommendation. Tang et al. (2020) introduced AMR\cite{tang2019adversarial}, which combines adversarial training with multimedia recommendation using visual features to enhance multimedia recommendation robustness. Paul et al. (2022) proposed ADCFA\cite{paul2022robust}, which combines dynamic CF with adversarial training to achieve robust multimedia recommendation.


While adversarial training can significantly enhance the robustness of recommender systems, it's important to note that controlling the intensity of adversarial noise during training can be challenging. This difficulty in controlling noise levels can, to some extent, reduce the model's ability to generalize effectively. Additionally, because adversarial training involves a minimax game, it strikes a balance between accuracy and robustness, which means that it optimizes for both, but not necessarily to the highest degree for either.

\subsubsection{Trust-aware Recommender}
In addition to model-based methods, trust information is also widely employed in robust algorithms for recommender systems. These approaches aim to identify trusted users or trusted items for recommending to users, thereby reducing the impact of malicious attacks.

Based on whether the models require external data beyond interaction data, trust-aware recommendation methods can be categorized into those that utilize external data and those that do not.
External data typically refers to social data, which records information about user interactions such as following and friend relationships. Some research leverages social data to mine trust information and then combines it with recommendation models to enhance robustness.

For example, Quan et al. (2023) designed the Preference-Guided Denoising Method with Social Relationships (GDMSR)\cite{quan2023robust}. GDMSR models confidence in social relationships and provides a denoised and richer social graph to enhance user preference learning. To mitigate noisy social relationships, GDMSR also incorporates a self-correcting curriculum learning module and an adaptive denoising strategy, both of which contribute to improved trust in high-degree samples. Wang et al. (2023) introduced the Denoised Self-Augmented Learning (DSL) paradigm\cite{wang2023denoised}. DSL starts with self-supervised denoising, preserving useful social relationships. It then employs a dual-view graph neural network to encode latent representations from user social and interaction graphs. A learnable similarity mapping function aligns these two views, achieving robust recommendations.

Methods that do not rely on external data typically mine information from existing data, evaluating score similarity, user trustworthiness, and other factors before combining them with recommendation models for the recommendation process.

For example, Jia et al. (2013) \cite{jia2013robust} assessed the credibility of user ratings based on item recommendation reliability, score similarity, and user trustworthiness. They combined a trust computation model with traditional CF methods, selecting a reliable set of neighbors to generate recommendations. Dongyan and Fuzhi (2014) \cite{jia2014robust} used entropy theory and density-based local outlier factors to calculate user suspicion levels. They proposed a multidimensional trust model to measure user trust from multiple perspectives, which was then combined with baseline estimation methods to form the Trusted Neighborhood Model. Gao et al. (2014) \cite{gao2014robust} proposed a user relationship-based method to enhance robustness. They constructed user models based on three types of relationships (interest similarity, score similarity, and trust), used density clustering to detect spam users, and introduced a detection-based item similarity measure within an item-based kNN CF framework. Experiments demonstrated its effectiveness in spam detection and robustness improvement. Yi and Zhang (2016) \cite{yi2016robust} introduced RRM-SUMMT, manually labeled suspicious users, and trained a relevance vector machine to determine suspicion levels. They then built a multidimensional trust model from rating data and combined MF with neighborhood models for recommendation.

\subsection{Hybrid Method and Others}
To simultaneously enhance the accuracy and robustness of recommender systems, some studies have combined detection and robust algorithms for recommendation.

For example, Yu et al. (2019)\cite{yu2019robust} employed k-means to detect anomalous users and integrated the detection results with Bayesian Probabilistic Matrix Factorization (BPMF) to achieve robust recommendations. Zhang et al. (2020) proposed GraphRfi\cite{zhang2020gcn}, which consists of both detection and recommendation components. In the fraud detection component, the probability of a user being identified as a fraudster automatically determines the contribution of their rating data in the recommendation component. The prediction error output in the recommendation component serves as an important feature in the fraud detection component. These two components mutually reinforce each other, resulting in robust and accurate recommendations. Wang et al. (2022)\cite{wang2022towards} employed three models for robustness training. In each training round, they used high-confidence predictions (consistent ratings) from any two models as auxiliary training data for the remaining model. This collaboration among the three models improved recommendation robustness.

\section{Non-adversarial Robustness}\label{sec:5}
\subsection{Definition and Taxonomy}\label{sec:5.1}
In addition to deliberate malicious attacks, there are other factors that can negatively impact the performance of recommender systems, such as user unintentional errors and unreasonable data distributions. The ability of a recommender system to withstand these factors and maintain its performance in non-adversarial situations is referred to as the system's non-adversarial robustness.

Based on research, we have identified three non-adversarial robustness factors that can influence recommendation performance. 
Data sparsity can lead to a decrease in model performance. When there is a lack of sufficient user interaction data, recommendation models may struggle to make accurate predictions.
Implicit feedback data often contains natural noise, including ``noisy positive examples'' (interactions that users made but didn't necessarily like) and ``noisy negative examples'' (cases where users didn't interact but might have liked). Handling this noise is crucial for robust recommendations.
Data imbalance can mislead traditional models that assign balanced class weights during the learning process. This can result in a reduction in recommendation accuracy, as the model may be biased towards the majority class.

In addressing the issue of non-adversarial robustness in recommender systems, this section explores and introduces corresponding solutions.

\subsection{Data Sparsity}
Due to the sparsity of user-item interaction matrices in recommender systems, where users typically rate only a small number of items, the limited number of ratings can make it challenging for models to learn effectively, thus impacting recommendation accuracy.



To learn more accurate representations on sparse datasets, one approach is to enhance data richness by introducing external content or using collaborative recommendation techniques to recommend items for users with sparse interactions.

For example, Li and She (2017) proposed a Bayesian generative model based on Collaborative Variational Autoencoders (CVAE)\cite{li2017collaborative} that integrates ratings and recommendation content in multimedia scenarios. It learns deep latent representations from content data in an unsupervised manner and captures implicit relationships between items and users of both content and ratings. Experiments showed that CVAE alleviates data sparsity and cold start problems. 
Chen et al. (2019)\cite{chen2019improved} improved the performance of recommendations in hash-based recommender systems by transforming locality sensitive hammering (LSH) into multiple probing LSH and integrating it into the auxiliary recommendation process, using collaborative recommendation principles to improve performance in cases of sparse interactions. 
Yang et al. (2021) propose an enhanced graph learning network for CF called EGLN \cite{Yang2021}. EGLN iterates between enhanced graph learning and node embedding learning, adding potential links absent in the input graph. By maximizing mutual information between local and global representations, it strengthens user-item relationships and improves performance on sparse data.
Yang et al. (2022)\cite{yang2022knowledge} combined knowledge graph data with contrastive learning to take advantage of the rich data in knowledge graphs to learn the true preferences of users with sparse interactions, thus improving the precision of the recommendation.
Similarly, Du et al. (2023)\cite{Du2023} proposed a metalearning framework based on knowledge graphs, MetaKG. Using a collaborative-aware meta-learner to capture user preferences and a knowledge-aware meta-learner to capture entity knowledge, MetaKG integrates an adaptive task scheduler to filter information tasks, ensuring recommendation performance in scenarios of data sparsity and cold starts.

\subsection{Natural Noise in Implicit Feedback}
Early work in recommender systems focused on explicit feedback tasks, such as training models to predict ratings using data like the ratings from MovieLens 1 to 5. However, explicit feedback suffers from data singularity and sparsity issues. Therefore, researchers began utilizing implicit feedback to build recommendation models for learning user preferences. Implicit feedback refers to user behaviors that can indicate user preferences, such as clicks, browsing, or dwell time. In specific work, implicit feedback is often represented in binary form.

However, there is a certain gap between the user preferences inferred from implicit feedback and the actual user preferences. For example, in an e-commerce recommendation scenario, a user may click on a product out of curiosity but has no intention of purchasing it, or they may not come across items of interest due to the low popularity of those items. Therefore, blindly fitting implicit feedback into a recommender system without considering natural noise can lead to a poor understanding of users' true preferences, ultimately harming the user experience and reducing recommendation performance.

Building upon this, the denoising of implicit feedback recommender systems holds significant importance. According to our research findings, we classify the denoising task of implicit feedback into two categories: sample selection methods and sample reweighting methods.

\subsubsection{Sample Selection Methods}
Sample selection emphasizes the use of well-designed sampling techniques to filter clean samples for learning user preferences. The recommendation performance of such methods relies heavily on the performance of the sampler.

For instance, Gantner et al. (2011) introduced the Bayesian Personalized Ranking framework WBPR\cite{gantner2012personalized}, applicable to scenarios with uneven sampling. The core idea of WBPR is that popular but non-interacted items are likely genuine negatives. Zhang et al. (2013) \cite{zhang2013optimizing} hypothesized that during training, any unobserved item should not rank higher than any observed positive item. Based on this, they proposed DNS, which dynamically selects negative samples from ranking lists generated by the current model. It uses an n-degree polynomial weighted function to select n+1 samples and compute sampling probabilities. Wang et al. (2021) \cite{wang2021implicit} presented an iterative re-ranking framework, IRBPR, to alleviate noise in implicit feedback. The re-ranking module applies self-training to dynamically generate pseudo-labels based on user preferences, which are then used to train downstream recommender modules. Experiments show IRBPR effectively mitigates noise. Zhu et al. (2022) \cite{zhu2022gain} noted that previous work lacked sufficient robustness to handle false negatives during training. To address this, they proposed GDNS, which employs an expected gain sampler to dynamically guide negative sample selection based on the expected preference gap between positive and negative items. This gain-adjusted sampler effectively identifies false negatives and reduces the risk of introducing them.

\subsubsection{Sample Reweighting Methods}
While sample selection methods can address the noise issue to some extent in implicit feedback recommender systems, they may unavoidably remove genuine samples due to the pursuit of high recall rates. Additionally, the performance of sample selection methods heavily relies on the data distribution.
To address the aforementioned challenges, a series of denoising methods based on sample reweighting have been proposed. Sample reweighting methods distinguish between clean and noisy samples based on sample loss values, assigning lower weights to noisy samples during model training\cite{gao2022self}.

For example, Chen et al. (2019)\cite{chen2019improving} introduced the rank-based implicit regularization method SRRMF, which penalizes excessive differences between unobserved samples to capture relationships between data. This approach makes the scores of unobserved interactions (negative samples) closer to each other, avoiding the use of undefined true ratings. Combining SRRMF with MF improved system performance. Wang et al. (2021)\cite{wang2021denoising} proposed the ADT framework, which assigns zero weights to samples with large losses using truncated BCE with dynamic thresholds. Wang et al. (2022)\cite{wang2022learning} observed that different models make similar predictions for clean samples and divergent predictions for noisy examples. They developed DeCA, a cross-model protocol for denoising, which minimizes the KL divergence between two models' parameterized user preferences and maximizes the likelihood of observed data given true user preferences. Gao et al. (2022)\cite{gao2022self} introduced SGDL, which collects memory interactions during the early training stages (the ``noise resistance period``) and uses them as denoising signals to guide the model's subsequent training (the ``noise-sensitive period``). SGDL automatically switches its learning phase from memory to self-guided learning and employs an adaptive denoising scheduler to select clean and informative memory data to enhance robustness. Wang et al. (2023)\cite{wang2023efficient} formulated recommendation denoising as a two-layer optimization problem called BOD. The inner optimization aims to derive effective recommendation models and guide weight determination, eliminating the need for prior knowledge. The outer optimization uses gradients from the inner optimization and considers the impact of previous weights to adjust them. Experiments demonstrate that BOD achieves good denoising results in implicit feedback, enhancing the performance of recommender systems.

\subsection{Data Imbalance}
Due to the inherent mechanics of recommender systems, the number of negative samples is often significantly greater than that of positive samples. Training models on imbalanced data can lead to a bias towards predicting the more frequent class, thereby reducing model performance. Moreover, a substantial amount of irrelevant data can also increase model training time.To address the issue of data imbalance, Yang et al.(2018)\cite{yang2018robust} proposed the ROMA recommendation model. ROMA combines cost-sensitive learning with constrained one-sided loss integrated into the joint objective function. This approach directs the model's focus more towards predicting positive instances, thereby enhancing model performance when dealing with imbalanced data.

\begin{table*}
\centering
\setlength{\abovecaptionskip}{-0.05cm} 
\setlength{\belowcaptionskip}{-0.2cm} 
  \caption{Explicit Feedback Dataset}
  \label{tab:ratingdata}
  \resizebox{\linewidth }{!}{
  \begin{tabular}{cccccccccc}

    \toprule
        Dataset&\#Users&\#Itmes&\#Ratings& Rating range& Sparsity&Avg Actions/User&Avg Actions/Item &Timestamp&Social data\\
    \midrule
        MovieLens-100K\footnotemark[1] &943 &1,682 &100,000 
& 1-5& 93.70\%& 106.04 
& 59.45 
&\checkmark & \\
 MovieLens-1M\footnotemark[1]& 6,040 & 3,706 & 1,000,209 
& 1-5& 95.53\%& 165.60 
& 269.89 
&\checkmark & \\
 MovieLens-10M\footnotemark[1]& 69,878 & 10,681 & 10,000,054 
& 1-5& 98.66\%& 143.11 
& 936.25 
&\checkmark & \\
 MovieLens-20M\footnotemark[1]& 138,493 & 27,278 & 20,000,263 
& 0.5-5.0& 99.47\%& 144.41 
& 733.20 
&\checkmark & \\
 Amazon Review 2014\footnotemark[2]& 20,980,000 & 9,350,000 & 82,830,000 
& 1-5& 99.99\%& 3.95 
& 8.86 
&\checkmark & \\
 Yelp\footnotemark[3]& 1,987,897 & 150,346 & 6,990,280 
& 1-5& 99.99\%& 3.52 
& 46.49 
&\checkmark & \checkmark \\
 Netflix\footnotemark[4]& 480,189 & 17,770 & 100,480,507 
& 1-5& 98.82\%& 209.25 
& 5654.50 
&\checkmark & \\
 Epinions\footnotemark[5]& 22,164 & 296,277 & 922,267 
& 1-5& 99.99\%& 41.61 
& 3.11 
&\checkmark & \checkmark \\
 FilmTrust\footnotemark[6]& 1,508 & 2,071 & 35,497 
& 0.5-4.0& 98.86\%& 23.54 
& 17.14 
&& \checkmark \\
 Ciao\footnotemark[7]& 10,877 & 103,935 & 269,197 
& 1-5& 99.98\%& 24.75 
& 2.59 
&\checkmark & \checkmark \\
 CiaoDVD\footnotemark[8]& 17,615 & 16,121 & 72,665 
& 1-5& 99.97\%& 4.13 
& 4.51 
&\checkmark & \checkmark \\
        Book-Crossing\footnotemark[9]&105,283 &340,556 &1,149,780 
& 0-10& 99.99\%& 10.92 
& 3.38 
&& \\
    \bottomrule
  \end{tabular}
  }
\end{table*}

\begin{table*}
\setlength{\abovecaptionskip}{-0.05cm} 
\setlength{\belowcaptionskip}{-0.2cm} 
\centering
\caption{Implicit Feedback Dataset}
\label{tab:implicitdata}
\begin{tabular}{ccccccccc}
\toprule
Dataset&\#Users& \#Itmes& \#Interactions&  Sparsity&Avg Actions/User&Avg Actions/Item &Timestamp& Social data\\
\midrule
Last.fm\footnotemark[10]& 1,892& 17,632& 92,834&   99.72\%&49.07 &5.27 &\checkmark & \checkmark \\
Pinterest\footnotemark[11]&  55,187& 9,916& 1,500,809&   99.73\%&27.19 &151.35 &&\\
Gowalla\footnotemark[12]&  107,092& 1,280,969& 6,442,892&   99.99\%&60.16 &5.03 &\checkmark &\checkmark \\
Adressa 2M Compact\footnotemark[13]& 15,514& 923& 2,717,915&   81.02\%&175.19 &2944.65 &\checkmark & \\
\bottomrule
\end{tabular}
\end{table*}

\section{Dataset for Robust Recommendation}\label{sec:6}
We investigated and compiled commonly used datasets for evaluating the robustness of recommender systems. These datasets were categorized into explicit feedback datasets and implicit feedback datasets based on whether they contain explicit user preference information. The statistical information can be found in Table~\ref{tab:ratingdata} and Table~\ref{tab:implicitdata}.

\subsection{Explicit Feedback Dataset}
MovieLens, as introduced by Harper and Konstan in their work \cite{harper2015movielens}, stands as one of the most classic and widely employed datasets in the realm of recommender systems, extensively utilized in both research and industrial applications. Originally introduced in 1998, MovieLens comprises user ratings for movies, represented in the form of basic triplets (userid, movieid, timestamp). Over the course of time, various versions of MovieLens datasets have been released, including 100k, 1M, 10M, 20M, and more, each varying in size and complexity. 

The Amazon Review dataset encompasses an extensive collection of customer reviews from the Amazon e-Commerce platform. This dataset includes information on customer reviews and ratings, relationships between products, timestamps, helpful votes, product images, prices, product categories, and sales data. Amazon has partitioned the dataset into multiple subdatasets based on various categories, such as Amazon Fashion, All Beauty, and Books, among others. Additionally, Amazon continues to update and release new versions of this dataset. As of now, there are three major versions available: Amazon-2013, Amazon-2014, and Amazon-2018, each providing a wealth of valuable data for research and analysis. 

The Yelp dataset comprises review data collected from the well-known American business review website, Yelp. This data set includes various categories of information, such as business data (including business id and address), review data (comprising stars, date, and text), user data (including user id, review count, and friends), as well as check-in data, tip data, and photo data.

The Netflix dataset is a movie rating dataset provided by the Netflix Prize competition. It is used to seek the best CF algorithms for predicting movie ratings. The fundamental data is structured as 4-tuples:  (user, movie, date of rating, rating) , and it also includes movie titles and release year information. Netflix has been extensively employed for evaluating rating prediction tasks.

The Epinions dataset \cite{tang2012etrust} is a rating dataset obtained from the product review website Epinions.com. A notable feature of the Epinions review website is the existence of trust networks among users. Therefore, the Epinions dataset includes data on trust relationships between users. The Epinions rating data encompasses product names, product categories, rating scores, timestamps of ratings, and the usefulness of the ratings.

The FilmTrust dataset \cite{10.5555/2540128.2540506}is a small-scale rating dataset obtained through web scraping from the FilmTrust website. This dataset also incorporates trust data between users. The rating data is structured as triplets (user-id, item-id, rating-value), while the trust data comprises triplets (user-id (trustor), user-id (trustee), trust-value).

The Ciao dataset \cite{tang2012etrust}is a rating dataset collected from the Ciao product review website. It includes both rating data and user trust data.

The CiaoDVD dataset \cite{guo2014etaf}is a dataset obtained by scraping web pages from the DVD category of the product review website dvd.ciao.co.uk. This dataset comprises movie ratings data (including userId, movieId, movie-categoryId, reviewId, movieRating, and reviewDate), review ratings data (including userId, reviewId, and helpfulness), and trust data (including trustorId, trusteeId, and trustRating).

The Book-Crossing Dataset \cite{ziegler2005improving} is an online book review dataset collected and released by a group of individuals, comprising book ratings, titles, and user ages.

\footnotetext[1]{\url{https://grouplens.org/datasets/movielens/}}
\footnotetext[2]{\url{https://cseweb.ucsd.edu/~jmcauley/datasets.html}}
\footnotetext[3]{\url{https://www.yelp.com/dataset}}
\footnotetext[4]{\url{https://www.kaggle.com/datasets/netflix-inc/netflix-prize-data/data}}
\footnotetext[5]{\url{https://www.cse.msu.edu/~tangjili/datasetcode/truststudy.htm}}
\footnotetext[6]{\url{https://guoguibing.github.io/librec/datasets.html}}
\footnotetext[7]{\url{https://www.kaggle.com/datasets/aravindaraman/ciao-data}}
\footnotetext[8]{\url{https://guoguibing.github.io/librec/datasets.html}}
\footnotetext[9]{\url{https://www.kaggle.com/datasets/somnambwl/bookcrossing-dataset}}
\footnotetext[10]{\url{https://grouplens.org/datasets/hetrec-2011/}}
\footnotetext[11]{\url{https://github.com/hexiangnan/neural_collaborative_filtering}}
\footnotetext[12]{\url{http://snap.stanford.edu/data/loc-gowalla.html}}
\footnotetext[13]{\url{https://reclab.idi.ntnu.no/dataset/}}

\subsection{Implicit Feedback Dataset}
The Last.fm dataset is derived from the online music platform Last.fm and includes over one billion listening events. It records the number of times users listen to music and provides specific information such as artist, album, tag, country, year of publication, and number of listeners.

The Pinterest dataset, originally constructed by X. Geng and others for content-based image recommendation as mentioned in the work \cite{geng2015learning}, is an implicit feedback dataset. In the Pinterest dataset, each interaction record indicates that a user has ``pinned'' a particular image to their board.

The Gowalla dataset \cite{cho2011friendship} is a location-based check-in dataset where users share their locations by checking-in. The check-in dataset records information about the latitude, longitude, check-in time, and location ID in tuples (user, check-in time, latitude, longitude, location id). Additionally, Gowalla provides information about the friendship network among its users.

The Adressa dataset \cite{gulla2017adressa} is a news dataset released through a collaborative effort between the Norwegian University of Science and Technology (NTNU) and Adressavisen, a local newspaper in Trondheim, Norway. The Adressa dataset records user interactions with news articles, capturing data on the number of clicks and the time users spend reading each article.

In addition, some studies use rating datasets to train implicit models, which require the conversion of rating data into implicit feedback data. Two common conversion strategies exist: one approach involves directly removing the ``rating'' column from the rating data, thus considering all user-item interactions as positive feedback; alternatively, a suitable rating threshold can be set, where interactions with ratings exceeding the threshold are considered positive feedback, while interactions with ratings below the threshold or absent are considered negative feedback.

\section{Evaluation Metrics}\label{sec:7}
Based on varying task requirements, we categorize recommender system evaluation metrics into three classes: rating prediction metrics, top-k recommendation metrics, and robustness metrics. In this subsection, we will provide a detailed overview of each category of metrics.
\subsection{Rating Prediction Metrics}
Predictive rating tasks constitute a classic regression challenge within the domain of recommender systems. In this context, two commonly employed evaluation metrics are the Mean Absolute Error (MAE) and the Root Mean Squared Error (RMSE).
$\mathcal{U}$ represents the set of users, $\mathcal{I}$ represents the set of items, $p_{u,i}$ denotes the predicted score for a user-item pair, and $r_{u,i}$ represents the true score for the user-item pair.

The MAE serves as a benchmark for assessing the average absolute error between predicted values and actual values, regardless of the direction of the errors. The specific formula for its computation is as follows:
\begin{equation} 
MAE=\frac{\sum_{u\in\mathcal{U},i\in\mathcal{I}}|p_{u,i}-r_{u,i}|}{|\mathcal{U}|}
\end{equation}
RMSE is used to measure the average difference between the predicted values of a statistical model and the actual (true) values. It is a measure of the standard deviation of the residuals, where the residuals represent the distances between the regression line and the data points. The specific formula for calculating RMSE is as follows:
\begin{equation} 
RMSE=\sqrt{\frac{\sum_{u\in\mathcal{U},i\in\mathcal{I}}(p_{u,i}-r_{u,i})^2}{|\mathcal{U}|-1}}
\end{equation}

\subsection{Top-k Recommendation Metrics}

Rating prediction indicates a recommendation model's ability to estimate user preferences, but solely relying on ratings for item ranking can introduce significant errors. Users typically seek lists of items rather than individual scores, and each user may have different criteria for rating items. For instance, a rating of 3 might be interpreted differently by different users. Moreover, data sparsity and cold-start problems make directly predicting user-item ratings challenging. Therefore, top-k recommender systems and their associated evaluation metrics are commonly utilized. These systems focus on generating a set of items likely to be appreciated by the user based on implicit interactions, rather than predicting ratings. Evaluation metrics for top-k recommendation fall into non-ranking and ranking categories. Non-ranking metrics like Precision, Recall, and F1 disregard item order, while ranking metrics such as MAP, MRR, and NDCG consider item order, making them suitable for scenarios where it matters.

$K$ represents the length of the item list in the top-k recommendations. $\mathcal{U}$ denotes the set of users, $R(u)$ represents the set of items recommended to user $u$ with $|R(u)|=K$, and $T(u)$ represents the set of items with which user $u$ has interacted in the test data set.

Precision is the most fundamental metric in top-k recommender systems. Measures the proportion of actual interactions that are correctly recommended in the final recommendation list. The specific formula for calculating RMSE is as follows:
\begin{equation}
\mathrm{Precision}@K = \frac {
\sum_{u \in \mathcal{U}}|R(u) \cap T(u)|
} {
\sum_{u \in \mathcal{U}}|R(u)|
}
\end{equation}

Another fundamental metric in top-k recommender systems is Recall. Measure the proportion of actual user-item interactions that are successfully recommended in the top k list. The formula for Recall is as follows:
\begin{equation}
\mathrm{Recall}@K = \frac {
\sum_{u \in \mathcal{U}}|R(u) \cap T(u)|
} {
\sum_{u \in \mathcal{U}}|T(u)|
}
\end{equation}

Certainly, because Precision and Recall are interrelated, we can use F1-score to evaluate the performance of top-k recommendations. The formula for F1-score is as follows:
\begin{equation}
\mathrm{F1} = \frac{
2 \times Precision \times Recall
}{
Precision + Recall
} 
\end{equation}
Additionally, HitRate is one of the most commonly used evaluation metrics in top-k recommendations. It represents the proportion of users for whom the recommendations were successful. The calculation formula is as follows:
\begin{equation}
\mathrm{HR}@K = \frac{
\mathrm{hits}
}{
\left \vert \mathcal{U} \right \vert
}
\end{equation}
,where 'hits' represents the number of users for whom recommendations were successful.

For the ranking metrics in top-k recommendations, one of the most commonly used is the Normalized Discounted Cumulative Gain (NDCG). NDCG originates in the field of information retrieval and measures the relevance between the output sequence and the true preferences. NDCG values fall within the range of [0, 1], where a higher value indicates better ranking quality. The calculation formula is as follows:
\begin{equation}
\mathrm{NDCG}@N = \frac{DCG@K} {IDCG@K} 
\end{equation}
, where $\mathrm{DCG}@K$ represents the Discounted Cumulative Gain, which is calculated using the following formula:
\begin{equation}
\mathrm{DCG}@K = \sum_{i =1}^K
\frac{
rel_i
}{
\log_2 (i + 1)
}
\end{equation}
, IDCG, or Ideal Discounted Cumulative Gain, represents the situation where the recommended items are ranked in descending order of their true gains in an ideal scenario. The calculation formula is as follows:
\begin{equation}
\mathrm{IDCG}@K = \sum_{i =1}^{\left \vert REL_K \right \vert}
\frac{
2^{rel_i} - 1
}{
\log_2 (i + 1) 
}
\end{equation}
, Where $REL_K$ represents the top K items ranked in descending order of gains, and $rel_i \in$ {0,1}.

The Mean Reciprocal Rank (MRR) is another commonly used ranking metric that places a high emphasis on the position of the first item in the list. MRR measures whether the recommended item for a user is prominently (near the top) of the list. The calculation formula is as follows:
\begin{equation}
MRR = \frac {1}{|\mathcal{U}|} \sum_{i=1}^{|\mathcal{U}|}\frac{1}{rank_i}
\end{equation}
, where $rank_i$ represents the position in the ranking of the first item in the recommended list that matches the ground-truth item for the i-th user.

Certainly, $MAP$ is another commonly used ranking metric in recommender systems. Represents the average precision of all users. The formula for calculating $MAP$ is as follows:
\begin{equation}
\mathrm{MAP}@K = \frac{1}{
\left \vert \mathcal{U} \right \vert
}
\sum_{u \in \mathcal{U}} \mathrm{AP}@K
\end{equation}
\begin{equation}
AP@K = \frac{
\sum_{n=1}^K
\mathrm{Precision@n\times rel(n)} 
}{
{K}
}
\end{equation}
where $rel(n)$ is an indicator function that denotes whether the n-th item has been adopted or not.

\subsection{Robustness Metrics}
When evaluating the adversarial robustness of recommender systems, some research directly compares the changes in various traditional evaluation metrics before and after shilling attacks. A smaller change in evaluation metrics indicates better robustness of the model. However, this evaluation method may not be very intuitive. Some studies \cite{wang2022towards,wang2021robust,yu2017novel,li2022causal,yu2019robust,fulan2018robust,noh2015auro,kaur2016shilling} use specialized robustness evaluation metrics to assess the system's ability to withstand shilling attacks or other scenarios. These metrics mainly include Predict Shift (PS), Rank Improvement (RI), Drop Rate (DR), and Failure Rate.

PS is used to evaluate the robustness of rating prediction models. The calculation formula is as follows:
\begin{equation}
PS=\frac{\sum_{u\in U}\left(\hat{r}_{ui}^{\prime}-\hat{r}_{ui}\right)}{|U|}
\end{equation}
, where $\hat{r}_{ui}$ represents the predicted score by the recommendation model for the user-item pair before the attack, and $\hat{r}_{ui}^{\prime}$ represents the predicted score by the recommendation model for the user-item pair after the attack.

The T-metric evaluates ranking-based performance specifically on the target items of the attack. A lower T-metric value indicates a lower attack success rate and thus stronger robustness of the model against shilling attacks. Common examples include T-NDCG and T-HR.

RI, proposed by Wu et al. (2021) \cite{wu2021fight}, is employed to assess the robustness of ranking models before and after an attack. The calculation formula is as follows:
\begin{equation}
RI=1-(HR_{defense}-HR_{origin})/(HR_{attack}-HR_{origin})
\end{equation}
, where $HR_{origin}$ represents the HitRate value under the original conditions, $HR_{attack}$ represents the HitRate value under attack conditions, and $HR_{defense}$ represents the HitRate value when a defense strategy is applied.

DR, introduced in (Wu et al., 2022) \cite{wu2022neural}, is used to assess the impact of data distribution on model performance. It is calculated as follows:
\begin{equation}
DR=\frac{P_I-P_N}{P_I}
\end{equation}
,where $P_I$ represents the performance of the model on an independently and identically distributed test set (typically higher), $P_N$ represents the performance on a non-independently and identically distributed test set (typically lower, which can be evaluated by partitioning the dataset into different sub-datasets). The performance metric $P$ can be any relevant evaluation metric such as Hit Ratio and NDCG, among others.

The Failure Rate represents the probability of recommendation failure, where recommendation failure can be defined as the situation where none of the items in the recommended list for user u are hits. The calculation formula is as follows:
\begin{equation}
\text{Failure Rate}=\frac{N_f}N\times100\%.
\end{equation}

\begin{table*}
\setlength{\abovecaptionskip}{-0.05cm} 
\setlength{\belowcaptionskip}{3.5cm} 
    \centering
        \caption{Existing Recommender System Libraries}

    \begin{tabular}{>{\centering\arraybackslash}m{1.5cm}ccm{10cm}>{\centering\arraybackslash}m{1cm}c} 
        \toprule
         Library&  Language&  Framework&  \multicolumn{1}{c}{Description}
&   Release time&  \#Star\footnotemark[9]\\ 
         \midrule
         LibRec\footnotemark[1]&  Java&  DL4J&  An open-source Java library designed for rating and ranking tasks in recommender systems, supporting rapid deployment on platforms like Hadoop and Spark.&  2014&  3.2k\\ 
         \midrule
         Recommenders\footnotemark[2]&  Python&  Tensorflow&  This repository encompasses illustrative examples and optimal methodologies for constructing recommender systems, presented in the form of Jupyter notebooks.&  2018&  18k\\ 
         \midrule
         EasyRec\footnotemark[3]&  Python&  Tensorflow&  It supports training across multiple platforms and data sources, allowing the exploration of arbitrary parameters for enhanced flexibility in model optimization.&  2020&  1.5k\\ 
         \midrule
         ReChorus\footnotemark[4]&  Python&  Pytorch&  A rapid evaluation framework tailored for top-k recommendation models, featuring concise code and bolstered by multi-threaded batch preparation capabilities.&  2020&  490\\ 
         \midrule
         RecBole\footnotemark[5]&  Python&  Pytorch&  It supports general, sequential, context-aware, and knowledge-based recommendation tasks, along with 40+ datasets and preprocessing tools.&  2020&  3.2k\\ 
         \midrule
         RecAd\footnotemark[6]&  Python&  Pytorch&  An open benchmark for evaluating recommender attacks and defenses, supporting robustness assessment of models against common attacks.&  2023&  21\\ 
         \midrule
         ARLib\footnotemark[7]&  Python&  Pytorch&  Focuses on attacks and defenses in top-k recommendation by adapting adversarial attack methods to implicit feedback.&  2023&  74\\ 
         \midrule
         ShillingREC\footnotemark[8]&  Python&  Pytorch&  The framework supports adversarial attacks on both implicit and explicit feedback data, enabling the evaluation of different categories of recommendation models based on the data type and the nature of the recommendation task.&  2024&  --\\
        \bottomrule
    \end{tabular}
    
    \label{tab:openreclib}
\end{table*}

\section{Adversarial Robustness Evaluation Library and Experiment}\label{sec:8}
In this chapter, we aim to fairly evaluate poisoning attack models and recommendation models under adversarial robustness scenarios to analyze the performance and applicability of different models. Therefore, we investigated open-source recommendation system frameworks, as shown in Table ~\ref{tab:openreclib}. Some open-source frameworks support the evaluation of poisoning attack and defense models, such as RecAd \cite{wangRecADUnifiedLibrary2023} and ARLib \cite{wangPoisoningAttacksRecommender2024}. However, these frameworks are not without limitations. RecAd only supports poisoning attacks on rating datasets, lacking adaptation for implicit data. In contrast, ARLib is designed for implicit data but forces poisoning attack methods for rating data to be adapted for implicit feedback data during evaluation. This setup is unfair to the former, as rating-based attack methods require generating specific fake ratings, while implicit data attack methods only need to generate fake interaction data.

\begin{figure}[h]
\setlength{\abovecaptionskip}{-0.1cm}

  \centering
  \includegraphics[width=\linewidth]{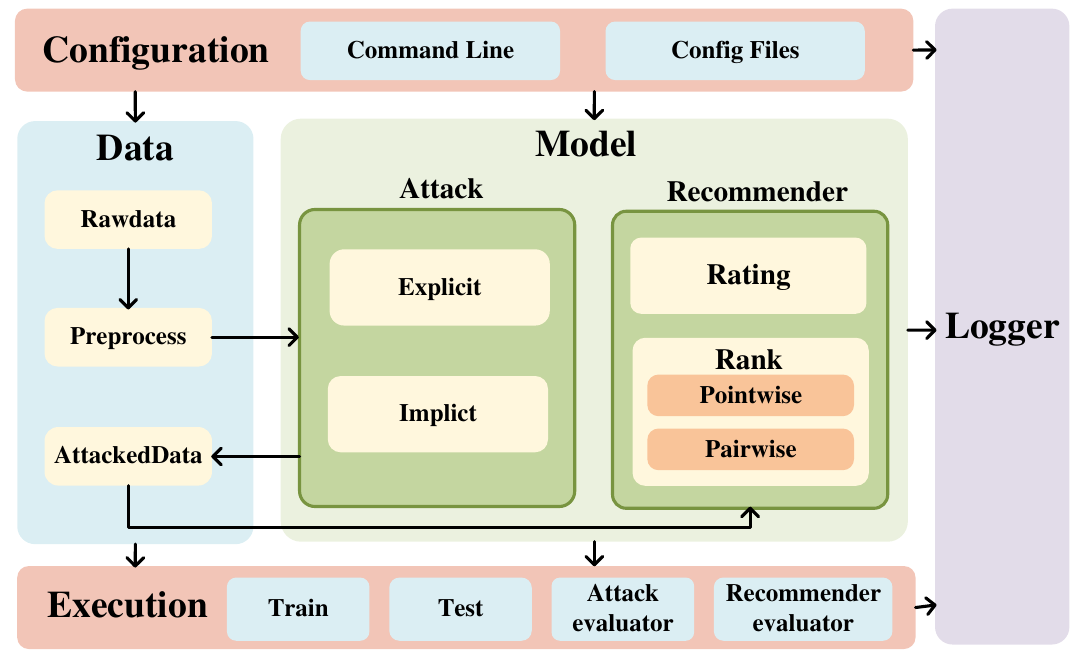}
  \caption{The overall framework of ShillingREC.}
  \label{fig:shillingrec}
    \vspace{-1 em}
\end{figure}

\subsection{The Library: ShillingREC}
To address the aforementioned issues, we propose ShillingREC, an adversarial robustness evaluation library that supports multiple attack modes and recommendation tasks. The overall framework of ShillingREC is illustrated in Figure \ref{fig:shillingrec}. ShillingREC enables poisoning attacks on both rating and implicit data formats. For rating data, it injects fake rating data and evaluates model performance using rating prediction tasks. For implicit feedback data, it injects fake interaction data and evaluates performance through top-k recommendation tasks. This design allows for a fair comparison of different poisoning attack models within their respective areas of strength. Additionally, ShillingREC supports various commonly used datasets, offering excellent scalability. It also provides efficient execution pipelines and comprehensive persistence strategies for data processing, poisoning attacks, and recommendation model training.

ShillingREC consists of five main components: Configuration Module, Data Module, Model Module, Execution Module, and Logger Module. 

\subsubsection{Configuration Module}
The Configuration Module utilizes YAML-format files to store configuration information. When ShillingREC is executed, the Configuration Module first retrieves top-level parameters from the ``antecedent.yaml'' file, such as the data name, attack model, and recommendation model. Subsequently, it loads specific parameters based on the data name or model information, among other factors.

\footnotetext[1]{\url{https://github.com/guoguibing/librec}}
\footnotetext[2]{\url{https://github.com/recommenders-team/recommenders}}
\footnotetext[3]{\url{https://github.com/alibaba/EasyRec}}
\footnotetext[4]{\url{https://github.com/THUwangcy/ReChorus/tree/master}}
\footnotetext[5]{\url{https://github.com/RUCAIBox/RecBole}}
\footnotetext[6]{\url{https://github.com/gusye1234/recad}}
\footnotetext[7]{\url{https://github.com/CoderWZW/ARLib?tab=readme-ov-file}}
\footnotetext[8]{\url{https://github.com/chengleileilei/ShillingREC}}
\footnotetext[9]{\#Star represents the number of stars for the project on the GitHub website, with data collection ending on April 21, 2024.}

\subsubsection{Data Module}
Ensuring data consistency is a core requirement of ShillingREC, thus the Data Module plays a pivotal role within the entire library. The primary functions of the Data Module include data loading, data preprocessing, and data persistence. Firstly, the Data Module supports loading data from raw data sources, requiring users only to ensure consistency in column names across datasets. Additionally, it facilitates parameterized data preprocessing, enabling swift execution of the same preprocessing strategy across different datasets. Lastly, the Data Module incorporates robust persistence strategies. Tailored to different preprocessing and attack models, it effectively stores AttackedData according to various parameters, facilitating rapid retrieval and preventing redundant computations.

\begin{table*}[]
\setlength{\abovecaptionskip}{-0.05cm} 
\setlength{\belowcaptionskip}{-0.2cm} 
\centering
\caption{Evaluation Results of Explicit Feedback Data}
\label{tab:explicit_attack}
\resizebox{\linewidth }{!}{
\begin{tabular}{cccccccccccccc}
\toprule
\multirow{2}{*}{Attack Model} & \multirow{2}{*}{Rec Model} & \multicolumn{3}{c}{ML-100K} & \multicolumn{3}{c}{ML-1M} & \multicolumn{3}{c}{CiaoDVD} & \multicolumn{3}{c}{Epinions} \\
\cmidrule(lr){3-5} \cmidrule(lr){6-8} \cmidrule(lr){9-11} \cmidrule(lr){12-14} &  & RMSE & MAE & PS & RMSE & MAE & PS & RMSE & MAE & PS & RMSE & MAE & PS \\
\midrule
\multirow{4}{*}{None} & MF & 1.1778 & 0.9823 & 0 & 1.1113 & 0.9276 & 0 & 1.5482 & 1.3238 & 0 & 1.3183 & 1.1490 & 0 \\
 & NeuMF & 0.9556 & 0.7520 & 0 & 0.8996 & 0.7071 & 0 & 0.9602 & 0.7290 & 0 & 1.0445 & 0.7840 & 0 \\
 & LightGCN & 0.9773 & 0.7678 & 0 & 0.8785 & 0.6894 & 0 & 1.2865 & 0.9809 & 0 & 1.1404 & 0.8822 & 0 \\
 & SGL & 0.9974 & 0.7820 & 0 & 0.8955 & 0.7015 & 0 & 1.3868 & 1.0554 & 0 & 1.2133 & 0.9405 & 0 \\
 \midrule
\multirow{4}{*}{Random Attack} & MF & 1.1693 & 0.9779 & 1.6996 & 1.1090 & 0.9269 & 1.2985 & 1.4541 & 1.2490 & 2.2385 & 1.3213 & 1.1569 & 0.6326 \\
 & NeuMF & 0.9577 & 0.7548 & 1.9365 & 0.8975 & 0.7069 & 1.7194 & 0.9553 & 0.7270 & 0.8949 & 1.0570 & 0.7939 & 1.1908 \\
 & LightGCN & 0.9529 & 0.7519 & 1.8517 & 0.8723 & 0.6855 & 1.3987 & 1.2108 & 0.9346 & 1.4879 & 1.1036 & 0.8617 & 2.2824 \\
 & SGL & 0.9667 & 0.7618 & 2.0922 & 1.0701 & 0.8482 & 2.4756 & 1.3301 & 1.0234 & 1.4936 & 1.1678 & 0.9130 & 2.6397 \\
 \midrule
\multirow{4}{*}{Average Attack} & MF & 1.1831 & 0.9864 & 1.8206 & 1.1129 & 0.9294 & 1.4259 & 1.4535 & 1.2483 & 2.2080 & 1.3172 & 1.1489 & 0.4278 \\
 & NeuMF & 0.9619 & 0.7567 & 2.2095 & 0.9009 & 0.7037 & 2.2377 & 0.9545 & 0.7245 & 0.8340 & 1.0558 & 0.7864 & 1.0411 \\
 & LightGCN & 0.9595 & 0.7568 & 3.0650 & 0.8729 & 0.6860 & 2.4352 & 1.2141 & 0.9365 & 1.4376 & 1.1006 & 0.8555 & 2.3379 \\
 & SGL & 0.9745 & 0.7674 & 3.3349 & 0.8814 & 0.6926 & 2.9289 & 1.0204 & 1.3297 & 1.5289 & 1.1627 & 0.9058 & 2.5788 \\
 \midrule
\multirow{4}{*}{LoveHate Attack} & MF & 1.1790 & 0.9829 & 0.5294 & 1.1111 & 0.9271 & 0.8449 & 1.5052 & 1.2866 & 0.5978 & 1.3198 & 1.1503 & 0.5983 \\
 & NeuMF & 0.9575 & 0.7495 & 4.1547 & 0.8895 & 0.6968 & 3.9182 & 0.9578 & 0.7292 & 1.8924 & 1.0496 & 0.7861 & 3.6649 \\
 & LightGCN & 0.9784 & 0.7689 & 3.3856 & 0.8750 & 0.6857 & 5.2998 & 1.3038 & 0.9811 & 2.4647 & 1.1202 & 0.8647 & 4.5518 \\
 & SGL & 0.9886 & 0.7743 & 5.1468 & 0.8885 & 0.6964 & 4.2893 & 1.4142 & 1.0684 & 2.3064 & 1.1945 & 0.9249 & 3.8280 \\
 \midrule
\multirow{4}{*}{AUSH} & MF & 1.1845 & 0.9875 & 1.7807 & 1.1113 & 0.9278 & 1.7045 & 1.4983 & 1.2795 & 2.2927 & 1.3163 & 1.1478 & 0.4278 \\
 & NeuMF & 0.9631 & 0.7577 & 2.3155 & 0.8912 & 0.6967 & 2.1448 & 0.9679 & 0.7335 & 0.8692 & 1.0597 & 0.7915 & 0.8401 \\
 & LightGCN & 0.9673 & 0.7586 & 5.8488 & 0.8808 & 0.6906 & 2.3850 & 1.2566 & 0.9558 & 1.9112 & 1.1722 & 0.8992 & 3.4358 \\
 & SGL & 0.9999 & 0.7835 & 3.9206 & 0.8979 & 0.7030 & 3.3751  & 1.3879 & 1.0542 & 2.3133 & 1.2271 & 0.9439 & 4.2808 \cr
 \bottomrule
\end{tabular}
}
\end{table*}

\begin{table*}[]
\setlength{\abovecaptionskip}{-0.05cm} 
\setlength{\belowcaptionskip}{-0.2cm} 
\centering
\caption{Evaluation Results of Implicit Feedback Data(K=50)}
\label{tab:implicit_attack}
\resizebox{\linewidth }{!}{
\begin{tabular}{cccccccccccccccccc}
\toprule
\multirow{2}{*}{\begin{tabular}[c]{@{}c@{}}Attack\\ Model\end{tabular}} & \multirow{2}{*}{\begin{tabular}[c]{@{}c@{}}Rec\\ Model\end{tabular}} & \multicolumn{4}{c}{Gowalla} & \multicolumn{4}{c}{LastFM} & \multicolumn{4}{c}{ML-1M} & \multicolumn{4}{c}{CiaoDVD} \\
 \cmidrule(lr){3-6} \cmidrule(lr){7-10} \cmidrule(lr){11-14} \cmidrule(lr){15-18} &  & NDCG & HR & T-NDCG & T-HR & NDCG & HR & T-NDCG & T-HR & NDCG & HR & T-NDCG & T-HR & NDCG & HR & T-NDCG & T-HR \\
 \midrule
\multirow{6}{*}{None} & MF & 0.0303 & 0.1491 & 0 & 0.2000 & 0.0461 & 0.4133 & 0.0002 & 0.0014 & 0.1346 & 0.8027 & 0 & 0 & 0.0359 & 0.2900 & 0.0114 & 0.0433 \\
 & NeuMF & 0.0668 & 0.3174 & 0.0003 & 0.0038 & 0.0676 & 0.5249 & 0 & 0 & 0.1575 & 0.8800 & 0 & 0 & 0.0938 & 0.5628 & 0 & 0 \\
 & LightGCN & 0.0821 & 0.3621 & 0.0001 & 0.0016 & 0.0674 & 0.5130 & 0 & 0 & 0.1612 & 0.8862 & 0 & 0 & 0.0874 & 0.5195 & 0.0044 & 0.0130 \\
 & CDAE & 0.0226 & 0.1553 & 0 & 0 & 0.0614 & 0.5151 & 0 & 0 & 0.1618 & 0.8966 & 0 & 0 & 0.0665 & 0.4199 & 0 & 0 \\
 & SGL & 0.0799 & 0.3599 & 0 & 0.0002 & 0.0557 & 0.4765 & 0.0005 & 0.0021 & 0.1609 & 0.8776 & 0 & 0 & 0.0999 & 0.5628 & 0.0192 & 0.0563 \\
 & SGDL & 0.0642 & 0.0002 & 0 & 0 & 0.0674 & 0.5207 & 0 & 0 & 0.1584& 0.8529& 0& 0& 0.0831 & 0.5043 & 0.0011 & 0.0043 \\
 \midrule
\multirow{6}{*}{\begin{tabular}[c]{@{}c@{}}Random\\ Attack\end{tabular}} & MF & 0.0278 & 0.1397 & 0.0146 & 0.0861 & 0.0384 & 0.3726 & 0.0027 & 0.0105 & 0.1368 & 0.8000 & 0 & 0 & 0.0415 & 0.3074 & 0.0196 & 0.0649 \\
 & NeuMF & 0.0616 & 0.2977 & 0.3026 & 0.5843 & 0.0577 & 0.5081 & 0.4606 & 1 & 0.1549 & 0.8662 & 0.0003 & 0.0018 & 0.0826 & 0.5498 & 0.9089 & 0.9307 \\
 & LightGCN & 0.0796 & 0.3556 & 0.0173 & 0.0780 & 0.0663 & 0.5109 & 0 & 0 & 0.1593 & 0.8764 & 0 & 0 & 0.0975 & 0.5649 & 0.0369 & 0.1147 \\
 & CDAE & 0.0184 & 0.1457 & 0.9977 & 1 & 0.0554 & 0.5039 & 0.7955 & 0.8561 & 0.1622 & 0.8995 & 0.0001 & 0.0017 & 0.0637 & 0.4091 & 0.2332 & 0.9978 \\
 & SGL & 0.0772 & 0.3462 & 0.0291 & 0.1663 & 0.0605 & 0.4940 & 0.0035 & 0.0119 & 0.1608 & 0.8751 & 0.0001 & 0.0003 & 0.0931 & 0.5693 & 0.0028 & 0.0130 \\
 & SGDL & 0.0704& 0.3433& 0.0701& 0.2398& 0.0668 & 0.5158 & 0 & 0 & 0.1587& 0.8596& 0& 0.0005& 0.0831 & 0.5152 & 0.0024 & 0.0087 \\ 
 \midrule
\multirow{6}{*}{\begin{tabular}[c]{@{}c@{}}Bandwagon\\ Attack\end{tabular}} & MF & 0.0278 & 0.1368 & 0.0385 & 0.1663 & 0.0361 & 0.3565 & 0.0058 & 0.0309 & 0.1339 & 0.7961 & 0.0001 & 0.0005 & 0.0403 & 0.3030 & 0.0169 & 0.0628 \\
 & NeuMF & 0.0624 & 0.2984 & 0.1650 & 0.4381 & 0.0615 & 0.5172 & 0.7368 & 1 & 0.1560 & 0.8672 & 0.0002 & 0.0017 & 0.0834 & 0.5519 & 0.6369 & 0.9199 \\
 & LightGCN & 0.0791 & 0.3543 & 0.0621 & 0.2202 & 0.0670 & 0.5235 & 0 & 0 & 0.1601 & 0.8813 & 0 & 0 & 0.0987 & 0.5801 & 0.0295 & 0.0779 \\
 & CDAE & 0.0190 & 0.1462 & 0.7748 & 1 & 0.0593 & 0.5193 & 0.0024 & 0.0189 & 0.1607 & 0.8963 & 0.0001 & 0.0008 & 0.0645 & 0.4134 & 0.2495 & 0.9978 \\
 & SGL & 0.0780 & 0.3534 & 0.0162 & 0.0700 & 0.0556 & 0.4814 & 0.0022 & 0.0077 & 0.1612 & 0.8754 & 0.0005 & 0.0039 & 0.0962 & 0.5563 & 0.0008 & 0.0043 \\
 & SGDL & 0.0693& 0.3424& 0.0761& 0.1851& 0.0675 & 0.5200 & 0 & 0 &  0.1591& 0.8576& 0& 0.0005& 0.0837 & 0.5130 & 0.0022 & 0.0087 \\
 \midrule
\multirow{6}{*}{AUSH} & MF & 0.0391 & 0.1969 & 0.0646 & 0.1558 & 0.0461 & 0.4456 & 0.5428 & 0.7733 & 0.1324 & 0.8022 & 0.1348 & 0.4569 & 0.0599 & 0.3939 & 0.0349 & 0.0801 \\
 & NeuMF & 0.0641 & 0.3100 & 0.1674 & 0.3684 & 0.0563 & 0.5165 & 0.9984 & 1 & 0.1579 & 0.8756 & 0.0193 & 0.0632 & 0.0824 & 0.5844 & 0.7730 & 0.8247 \\
 & LightGCN & 0.0802 & 0.3596 & 0.1634 & 0.3040 & 0.0565 & 0.5130 & 0.9977 & 0.9993 & 0.1616 & 0.8872 & 0.0006 & 0.0017 & 0.0874 & 0.5260 & 0.0077 & 0.0260 \\
 & CDAE & 0.0178 & 0.1426 & 0.9998 & 1 & 0.0544 & 0.5144 & 0.9296 & 0.9544 & 0.1599 & 0.8985 & 0.0005 & 0.0025 & 0.0636 & 0.4048 & 0.2016 & 0.9957 \\
 & SGL & 0.0797 & 0.3578 & 0.0494 & 0.1207 & 0.0532 & 0.4709 & 0.6427 & 0.8182 & 0.1610 & 0.8734 & 0.0004 & 0.0013 & 0.0963 & 0.5844 & 0.0041 & 0.0195 \\
 & SGDL & 0.0693& 0.3487& 0.1420& 0.2776& 0.0556 & 0.5039 & 0.9984 & 1 & 0.1595& 0.8543& 0.0033& 0.0183& 0.0820 & 0.5087 & 0.0077 & 0.0325 \\
 \midrule
\multirow{6}{*}{RevAdv} & MF & 0.0349 & 0.1676 & 0.0393 & 0.1632 & 0.0436 & 0.4126 & 0.2504 & 0.6912 & 0.1361 & 0.7926 & 0 & 0.0002 & 0.0180 & 0.1883 & 0.0005 & 0.0022 \\
 & NeuMF & 0.0654 & 0.3100 & 0.1672 & 0.5387 & 0.0616 & 0.5123 & 0.4192 & 1 & 0.1573 & 0.8754 & 0.0014 & 0.0097 & 0.0967 & 0.5758 & 0.2764 & 0.7208 \\
 & LightGCN & 0.0820 & 0.3637 & 0.0491 & 0.1690 & 0.0615 & 0.5102 & 0.7367 & 1 & 0.1607 & 0.8897 & 0.0021 & 0.0171 & 0.0851 & 0.5216 & 0.1734 & 0.6883 \\
 & CDAE & 0.0729 & 0.3364 & 0.2083 & 0.6596 & 0.0595 & 0.5060 & 0.1476 & 0.7895 & 0.1596 & 0.8983 & 0.0105 & 0.1032 & 0.0649 & 0.4156 & 0.1952 & 0.9935 \\
 & SGL & 0.0796 & 0.3612 & 0.0150 & 0.0773 & 0.0572 & 0.4772 & 0.1835 & 0.7460 & 0.1609 & 0.8798 & 0.0024 & 0.0173 & 0.0996 & 0.5801 & 0.0847 & 0.3506 \\
 & SGDL & 0.0699& 0.3438& 0.0279& 0.1477& 0.0635 & 0.4989 & 0.6017 & 1 & 0.1596& 0.8571& 0& 0& 0.0829 & 0.5130 & 0.1742 & 0.7100 \\
 \bottomrule
\end{tabular}
}
\end{table*}

\subsubsection{Model Module}
Model Module serves as the core component of ShillingREC, comprising the Attack Model and Rec Model. Within the Attack Model segment, ShillingREC categorizes shilling attack models into those tailored for explicit feedback data and those for implicit feedback data. Specifically, for explicit feedback data, the shilling attack model generates fake data in the form of (user-item-rating), while for implicit feedback data, it generates fake data in the form of (user-item). In the Rec Model segment, ShillingREC supports various common recommendation models such as MF\cite{koren2009matrix}, NeuMF\cite{He2017}, LightGCN\cite{heLightGCNSimplifyingPowering2020}, and SGL\cite{wu2021self}, with the ability to easily incorporate additional models. Furthermore, ShillingREC adapts different recommendation models to two tasks: rating prediction and top-k ranking, aiming to achieve a relatively fair evaluation in various experimental settings.

\subsubsection{Execution Module}
Execution Module is responsible for the operational logic of ShillingREC. Specifically, it automatically conducts tasks such as loading data and models, training and testing shilling attack models and recommendation models based on parameters provided by the Configuration Module, as well as model evaluation. The Execution Module exhibits strong flexibility: during the model training phase, it can employ different training strategies based on parameter settings, such as utilizing pointwise or pairwise strategies for ranking tasks; it can also employ various sampling strategies for implicit feedback tasks.

\subsubsection{Logger Module}
Logger Module serves as an auxiliary component of ShillingREC, dedicated to recording and preserving log information. Throughout each execution process of ShillingREC, the Logger Module diligently captures various types of information including execution parameters, data statistics, and model outputs. It dynamically configures appropriate log paths based on the data set and the experimental model information, ensuring the local persistence of the logged data.

\subsection{Experiment}
This section presents an evaluation of the attack performance of several typical adversarial attack models using ShillingREC in scenarios involving explicit data attacks or implicit feedback data attacks. 


\subsubsection{Explicit Feedback Data}
To evaluate the performance and robustness of recommendation models under shilling attacks in explicit feedback scenarios, we conduct rating prediction experiments on four widely used datasets: ML-100K, ML-1M, CiaoDVD, and Epinions. The evaluated recommendation models include MF, NeuMF, LightGCN, and SGL, covering  both classic and state-of-the-art methods. The attack models include Random Attack, Average Attack, Love/Hate Attack—widely adopted classical heuristic-based methods—and AUSH, a representative GAN-based attack. All attacks are configured with a consistent setup: the attacker size is set to 20\% of the training users; target items are selected from unpopular items, with the number of target items set to 0.5\% of the attacker size; and the number of filler items equals the average number of ratings per user in the dataset. We use MAE and RMSE to evaluate recommendation performance, and PS (Prediction Shift) to evaluate robustness. Results are presented in Table \ref{tab:explicit_attack}.

\subsubsection{Implicit Feedback Data}
To evaluate the performance and robustness of recommendation models under shilling attacks in implicit feedback scenarios, we conduct Top-K recommendation experiments on four widely used datasets: Gowalla, LastFM, ML-1M, and CiaoDVD. Gowalla and LastFM are native implicit feedback datasets, while ML-1M and CiaoDVD are converted from explicit to implicit feedback by applying a rating threshold. The evaluated recommendation models include MF, NeuMF, LightGCN, CDAE, SGL, and SGDL, covering both classic and state-of-the-art methods. The attack models include Random Attack and Bandwagon Attack—widely adopted classical heuristic-based methods—AUSH, a representative GAN-based attack, and RevAdv, a representative optimization-based attack. The attack configurations follow the same setup as in the explicit feedback experiments. 
We use NDCG and Hit Ratio to evaluate recommendation performance, and T-NDCG and T-HR to evaluate robustness, with all evaluation metrics computed under the Top-K recommendation setting, where K=50. Results are presented in Table~\ref{tab:implicit_attack}.

\subsubsection{Experimental Analysis}
First, the effectiveness of shilling attacks is sensitive to dataset size. Generally, attacks are more successful on small-scale datasets. For example, in explicit feedback scenarios, attack models achieve higher PS scores on the ML-100K dataset than on the larger ML-1M. This is because user interactions are sparser in smaller datasets, making models more vulnerable to the biased signals from fake users. In contrast, large datasets contain richer and more diverse user behaviors, where real data dominates and weakens the impact of injected fake data.

Second, in implicit feedback scenarios, models using unsupervised signals—such as SGL—show stronger robustness than supervised ones. Under AUSH attacks, SGL consistently outperforms LightGCN across all datasets in robustness metrics (T-NDCG and T-HR). This is due to the filtering effect of unsupervised contrastive learning, which enforces consistency and improves model stability and generalization under shilling attacks.

Lastly, regarding attack methods, optimization-based and model-based attacks (e.g., AUSH and RevAdv) generally outperform heuristic ones like Random Attack and Bandwagon Attack. On the LastFM dataset, for instance, AUSH consistently achieves better attack results across all models. This is because such methods don’t rely on fixed rules but generate more realistic and effective fake behaviors, leading to stronger manipulation.

\section{Future Research Direction and Open Issues}\label{sec:9}
From a data perspective, research on adversarial robustness in recommendation algorithms lacks real-world data of fraudulent users. Currently, academic studies are based on artificially simulated fake user data. This setup cannot guarantee the effectiveness of robust recommendation algorithms developed in actual production environments. Furthermore, due to limitations imposed by privacy concerns and platform regulations, the features of fake users generated by the attack model are restricted. Currently, it can only simulate user-item-rating triplet data and cannot replicate the complex features present in real-world scenarios. Moreover, treating adversarial attacks as an upstream task in the adversarial robustness of recommender systems may to some extent constrain the research on robust recommendation algorithms.

From a methodological point of view, there exists a trade-off between the effectiveness and efficiency of adversarial robust attack models. Generally, the more knowledge the attack model possesses about the recommender system, the better the attack's effectiveness; however, this often results in slower generation of fake data. In current applications, attacks are typically performed as offline tasks with low real-time requirements. Consequently, most existing research has focused primarily on the effectiveness of attack models, with limited attention given to the timeliness of adversarial attacks. 
In fraud detection models, aiming for a high recall rate can lead to sample misclassification, which may negatively impact subsequent recommendation model training. Joint training of fraud detection and recommendation tasks is a promising approach, as explored in works like GraphRfi \cite{zhang2020gcn} and PDR \cite{laiAdversariallyRobustRecommendation2023}. 
Similarly, in robust algorithms for adversarial defense strategies, performance and efficiency also present a trade-off. For example, methods based on contrastive learning and unsupervised strategies may introduce additional parameters, reducing the speed of inference. Given that recommendation models require rapid online inference, it is essential to achieve better robustness within acceptable timeliness limits. Therefore, researching high-timeliness robust recommendation algorithms is a critical direction.

From the perspective of model generalization, adversarial robustness largely depends on the design of the attack model, the scale of the attack, and the selection of attack targets. Most existing defense methods are tailored to specific attack strategies or base model designs, and currently, there is no sufficiently robust defense strategy that can protect various types of recommendation models against all forms of attack. Regarding non-adversarial robustness, current research primarily focuses on denoising techniques for natural noise. However, these methods are often limited to narrow scenarios and fail to address issues like data imbalance, sampling bias, and natural noise simultaneously. Thus, developing a general-purpose robust recommendation algorithm is a crucial research focus in recommendation system robustness.

\section{Conclusion}\label{sec:10}
In this survey, we provide a comprehensive overview of robustness in recommender systems. We summarize and categorize a large body of existing robustness methods into two main types: adversarial robustness and non-adversarial robustness. In the section on adversarial robustness of recommender systems, we delve into attack and defense perspectives, detailing the classification methods and basic principles of shilling attacks and defenses, along with an introduction to classical methods. In the non-adversarial robustness section, we explore the causes and corresponding methods for non-adversarial robustness issues stemming from data sparsity, natural noise, and data imbalance. Additionally, we present a comprehensive analysis and introduction of commonly used datasets and evaluation metrics in robustness research of recommender systems. To ensure fair and reasonable evaluation of existing shilling attack and defense methods in recommender systems, we propose an adversarial robustness evaluation library–ShillingREC and we experimentally assess several attack and recommendation models within this library. Finally, we analyze potential issues and future research directions in the field of recommender system robustness.

\section*{Acknowledgments}
This work was supported in part by the National Key Research and Development Program of China under Grant (2023YFC3310700),  the National Natural Science Foundation of China (62202041), the Fundamental Research Funds for the Central Universities (2023JBMC057).

\bibliographystyle{IEEEtran}
\bibliography{IEEEabrv,reference}

\begin{thebibliography}{10}
\providecommand{\url}[1]{#1}
\csname url@samestyle\endcsname
\providecommand{\newblock}{\relax}
\providecommand{\bibinfo}[2]{#2}
\providecommand{\BIBentrySTDinterwordspacing}{\spaceskip=0pt\relax}
\providecommand{\BIBentryALTinterwordstretchfactor}{4}
\providecommand{\BIBentryALTinterwordspacing}{\spaceskip=\fontdimen2\font plus
\BIBentryALTinterwordstretchfactor\fontdimen3\font minus \fontdimen4\font\relax}
\providecommand{\BIBforeignlanguage}[2]{{%
\expandafter\ifx\csname l@#1\endcsname\relax
\typeout{** WARNING: IEEEtran.bst: No hyphenation pattern has been}%
\typeout{** loaded for the language `#1'. Using the pattern for}%
\typeout{** the default language instead.}%
\else
\language=\csname l@#1\endcsname
\fi
#2}}
\providecommand{\BIBdecl}{\relax}
\BIBdecl

\bibitem{kaya2023robustness}
T.~T. Kaya and C.~Kaleli, ``Robustness analysis of multi-criteria top-n collaborative recommender system,'' \emph{Arabian Journal for Science and Engineering}, vol.~48, no.~8, pp. 10\,189--10\,212, 2023.

\bibitem{li2017collaborative}
X.~Li and J.~She, ``Collaborative variational autoencoder for recommender systems,'' in \emph{Proceedings of the 23rd ACM SIGKDD international conference on knowledge discovery and data mining}, 2017, pp. 305--314.

\bibitem{yang2022knowledge}
Y.~Yang, C.~Huang, L.~Xia, and C.~Li, ``Knowledge graph contrastive learning for recommendation,'' in \emph{Proceedings of the 45th International ACM SIGIR Conference on Research and Development in Information Retrieval}, 2022, pp. 1434--1443.

\bibitem{chen2019improved}
X.~Chen, H.~Liu, and D.~Yang, ``Improved lsh for privacy-aware and robust recommender system with sparse data in edge environment,'' \emph{EURASIP Journal on Wireless Communications and Networking}, vol. 2019, pp. 1--11, 2019.

\bibitem{deldjoo2021survey}
Y.~Deldjoo, T.~D. Noia, and F.~A. Merra, ``A survey on adversarial recommender systems: from attack/defense strategies to generative adversarial networks,'' \emph{ACM Computing Surveys (CSUR)}, vol.~54, no.~2, pp. 1--38, 2021.

\bibitem{JiShouling_2022}
J.~Shouling, D.~Tianyu, D.~Shuiguang, C.~Peng, S.~Jie, Y.~Min, and L.~Bo, ``A review of research on robustness of deep learning models,'' \emph{Journal of Computer Science}, vol.~45, no.~1, pp. 190--206, 2022.

\bibitem{si2020shilling}
M.~Si and Q.~Li, ``Shilling attacks against collaborative recommender systems: a review,'' \emph{Artificial Intelligence Review}, vol.~53, pp. 291--319, 2020.

\bibitem{ge2022survey}
Y.~Ge, S.~Liu, Z.~Fu, J.~Tan, Z.~Li, S.~Xu, Y.~Li, Y.~Xian, and Y.~Zhang, ``A survey on trustworthy recommender systems,'' \emph{arXiv preprint arXiv:2207.12515}, 2022.

\bibitem{fan2022comprehensive}
W.~Fan, X.~Zhao, X.~Chen, J.~Su, J.~Gao, L.~Wang, Q.~Liu, Y.~Wang, H.~Xu, L.~Chen \emph{et~al.}, ``A comprehensive survey on trustworthy recommender systems,'' \emph{arXiv preprint arXiv:2209.10117}, 2022.

\bibitem{OMahonyb}
M.~P. O'Mahony, N.~J. Hurley, and G.~C. Silvestre, ``Promoting recommendations: An attack on collaborative filtering,'' in \emph{Database and Expert Systems Applications}.\hskip 1em plus 0.5em minus 0.4em\relax Springer, 2002, pp. 494--503.

\bibitem{OMahonya}
M.~P. O'Mahony, ``Towards robust and efficient automated collaborative filtering,'' 2004.

\bibitem{liDataPoisoningAttacks2016}
B.~Li, Y.~Wang, A.~Singh, and Y.~Vorobeychik, ``Data poisoning attacks on factorization-based collaborative filtering,'' in \emph{Advances in Neural Information Processing Systems}, vol.~29.\hskip 1em plus 0.5em minus 0.4em\relax Curran Associates, Inc., 2016.

\bibitem{fangPoisoningAttacksGraphBased2018}
M.~Fang, G.~Yang, N.~Z. Gong, and J.~Liu, ``Poisoning attacks to graph-based recommender systems,'' in \emph{Proceedings of the 34th Annual Computer Security Applications Conference}.\hskip 1em plus 0.5em minus 0.4em\relax Association for Computing Machinery, 2018, pp. 381--392.

\bibitem{wuTripleAdversarialLearning2021}
C.~Wu, D.~Lian, Y.~Ge, Z.~Zhu, and E.~Chen, ``Triple adversarial learning for influence based poisoning attack in recommender systems,'' in \emph{Proceedings of the 27th ACM SIGKDD Conference on Knowledge Discovery \& Data Mining}.\hskip 1em plus 0.5em minus 0.4em\relax Association for Computing Machinery, 2021, pp. 1830--1840.

\bibitem{zhangPracticalDataPoisoning2020}
H.~Zhang, Y.~Li, B.~Ding, and J.~Gao, ``Practical data poisoning attack against next-item recommendation,'' in \emph{Proceedings of The Web Conference 2020}.\hskip 1em plus 0.5em minus 0.4em\relax Association for Computing Machinery, 2020, pp. 2458--2464.

\bibitem{songPoisonRecAdaptiveData2020}
J.~Song, Z.~Li, Z.~Hu, Y.~Wu, Z.~Li, J.~Li, and J.~Gao, ``Poisonrec: An adaptive data poisoning framework for attacking black-box recommender systems,'' in \emph{2020 IEEE 36th International Conference on Data Engineering (ICDE)}, 2020, pp. 157--168.

\bibitem{lamShillingRECommenderSystems2004}
S.~K. Lam and J.~Riedl, ``Shilling recommender systems for fun and profit,'' in \emph{Proceedings of the 13th International Conference on World Wide Web}.\hskip 1em plus 0.5em minus 0.4em\relax Association for Computing Machinery, 2004, pp. 393--402.

\bibitem{mobasherTrustworthyRecommenderSystems2007}
B.~Mobasher, R.~Burke, R.~Bhaumik, and C.~Williams, ``Toward trustworthy recommender systems: An analysis of attack models and algorithm robustness,'' \emph{ACM Transactions on Internet Technology}, vol.~7, no.~4, pp. 23--es, 2007.

\bibitem{burkeSegmentbasedInjectionAttacks2005}
R.~Burke, B.~Mobasher, R.~Bhaumik, and C.~Williams, ``Segment-based injection attacks against collaborative filtering recommender systems,'' in \emph{Fifth IEEE International Conference on Data Mining (ICDM'05)}, 2005, pp. 4 pp.--.

\bibitem{fangInfluenceFunctionBased2020}
M.~Fang, N.~Z. Gong, and J.~Liu, ``Influence function based data poisoning attacks to top-n recommender systems,'' in \emph{Proceedings of The Web Conference 2020}.\hskip 1em plus 0.5em minus 0.4em\relax Association for Computing Machinery, 2020, pp. 3019--3025.

\bibitem{Huang2021}
H.~Huang, J.~Mu, N.~Z. Gong, Q.~Li, B.~Liu, and M.~Xu, ``Data poisoning attacks to deep learning based recommender systems,'' in \emph{Proceedings 2021 Network and Distributed System Security Symposium}, 2021.

\bibitem{wuInfluenceDrivenDataPoisoning2023}
C.~Wu, D.~Lian, Y.~Ge, Z.~Zhu, and E.~Chen, ``Influence-driven data poisoning for robust recommender systems,'' \emph{IEEE TRANSACTIONS ON PATTERN ANALYSIS AND MACHINE INTELLIGENCE}, vol.~45, no.~10, 2023.

\bibitem{tangRevisitingAdversariallyLearned2020}
J.~Tang, H.~Wen, and K.~Wang, ``Revisiting adversarially learned injection attacks against recommender systems,'' in \emph{Fourteenth ACM Conference on Recommender Systems}, 2020, pp. 318--327.

\bibitem{zhangDataPoisoningAttack2021}
H.~Zhang, C.~Tian, Y.~Li, L.~Su, N.~Yang, W.~X. Zhao, and J.~Gao, ``Data poisoning attack against recommender system using incomplete and perturbed data,'' in \emph{Proceedings of the 27th ACM SIGKDD Conference on Knowledge Discovery \& Data Mining}.\hskip 1em plus 0.5em minus 0.4em\relax Association for Computing Machinery, 2021, pp. 2154--2164.

\bibitem{goodfellow2014generative}
I.~Goodfellow, J.~Pouget-Abadie, M.~Mirza, B.~Xu, D.~Warde-Farley, S.~Ozair, A.~Courville, and Y.~Bengio, ``Generative adversarial nets,'' \emph{Advances in neural information processing systems}, vol.~27, 2014.

\bibitem{christakopoulouAdversarialAttacksOblivious2019}
K.~Christakopoulou and A.~Banerjee, ``Adversarial attacks on an oblivious recommender,'' in \emph{Proceedings of the 13th ACM Conference on Recommender Systems}.\hskip 1em plus 0.5em minus 0.4em\relax Association for Computing Machinery, 2019, pp. 322--330.

\bibitem{linAttackingRecommenderSystems2020}
C.~Lin, S.~Chen, H.~Li, Y.~Xiao, L.~Li, and Q.~Yang, ``Attacking recommender systems with augmented user profiles,'' in \emph{Proceedings of the 29th ACM International Conference on Information \& Knowledge Management}.\hskip 1em plus 0.5em minus 0.4em\relax ACM, 2020, pp. 855--864.

\bibitem{zhangAttackingRecommenderSystems2021}
X.~Zhang, J.~Chen, R.~Zhang, C.~Wang, and L.~Liu, ``Attacking recommender systems with plausible profile,'' \emph{IEEE Transactions on Information Forensics and Security}, vol.~16, pp. 4788--4800, 2021.

\bibitem{wuReadyEmergingThreats2021}
F.~Wu, M.~Gao, J.~Yu, Z.~Wang, K.~Liu, and X.~Wang, ``Ready for emerging threats to recommender systems? a graph convolution-based generative shilling attack,'' \emph{Information Sciences}, vol. 578, pp. 683--701, 2021.

\bibitem{linShillingBlackBoxRecommender2022}
C.~Lin, S.~Chen, M.~Zeng, S.~Zhang, M.~Gao, and H.~Li, ``Shilling black-box recommender systems by learning to generate fake user profiles,'' \emph{IEEE Transactions on Neural Networks and Learning Systems}, pp. 1--15, 2022.

\bibitem{fanAttackingBlackboxRecommendations2021}
W.~Fan, T.~Derr, X.~Zhao, Y.~Ma, H.~Liu, J.~Wang, J.~Tang, and Q.~Li, ``Attacking black-box recommendations via copying cross-domain user profiles,'' in \emph{2021 IEEE 37th International Conference on Data Engineering (ICDE)}.\hskip 1em plus 0.5em minus 0.4em\relax IEEE, 2021, pp. 1583--1594.

\bibitem{zhou2016svm}
W.~Zhou, J.~Wen, Q.~Xiong, M.~Gao, and J.~Zeng, ``Svm-tia a shilling attack detection method based on svm and target item analysis in recommender systems,'' \emph{Neurocomputing}, vol. 210, pp. 197--205, 2016.

\bibitem{tong2018shilling}
C.~Tong, X.~Yin, J.~Li, T.~Zhu, R.~Lv, L.~Sun, and J.~J. Rodrigues, ``A shilling attack detector based on convolutional neural network for collaborative recommender system in social aware network,'' \emph{The Computer Journal}, vol.~61, no.~7, pp. 949--958, 2018.

\bibitem{ebrahimian2020detecting}
M.~Ebrahimian and R.~Kashef, ``Detecting shilling attacks using hybrid deep learning models,'' \emph{Symmetry}, vol.~12, no.~11, p. 1805, 2020.

\bibitem{mehta2009unsupervised}
B.~Mehta and W.~Nejdl, ``Unsupervised strategies for shilling detection and robust collaborative filtering,'' \emph{User Modeling and User-Adapted Interaction}, vol.~19, pp. 65--97, 2009.

\bibitem{zhang2015catch}
Y.~Zhang, Y.~Tan, M.~Zhang, Y.~Liu, T.-S. Chua, and S.~Ma, ``Catch the black sheep: unified framework for shilling attack detection based on fraudulent action propagation,'' in \emph{Twenty-fourth international joint conference on artificial intelligence}, 2015.

\bibitem{hao2021unsupervised}
Y.~Hao and F.~Zhang, ``An unsupervised detection method for shilling attacks based on deep learning and community detection,'' \emph{Soft Computing}, vol.~25, no.~1, pp. 477--494, 2021.

\bibitem{wu2012hysad}
Z.~Wu, J.~Wu, J.~Cao, and D.~Tao, ``Hysad: A semi-supervised hybrid shilling attack detector for trustworthy product recommendation,'' in \emph{Proceedings of the 18th ACM SIGKDD international conference on Knowledge discovery and data mining}, 2012, pp. 985--993.

\bibitem{zhou2021semi}
Q.~Zhou and L.~Duan, ``Semi-supervised recommendation attack detection based on co-forest,'' \emph{Computers \& Security}, vol. 109, p. 102390, 2021.

\bibitem{hao2023detection}
Y.~Hao, G.~Meng, J.~Wang, and C.~Zong, ``A detection method for hybrid attacks in recommender systems,'' \emph{Information Systems}, vol. 114, p. 102154, 2023.

\bibitem{cheng2010robust}
Z.~Cheng and N.~Hurley, ``Robust collaborative recommendation by least trimmed squares matrix factorization,'' in \emph{2010 22nd IEEE International Conference on Tools with Artificial Intelligence}, vol.~2.\hskip 1em plus 0.5em minus 0.4em\relax IEEE, 2010, pp. 105--112.

\bibitem{yi2014robust}
H.~Yi, F.~Zhang, and J.~Lan, ``A robust collaborative recommendation algorithm based on k-distance and tukey m-estimator,'' \emph{China Communications}, vol.~11, no.~9, pp. 112--123, 2014.

\bibitem{yu2017novel}
H.~Yu, R.~Gao, K.~Wang, and F.~Zhang, ``A novel robust recommendation method based on kernel matrix factorization,'' \emph{Journal of Intelligent \& Fuzzy Systems}, vol.~32, no.~3, pp. 2101--2109, 2017.

\bibitem{fadaee2018chiron}
S.~S. Fadaee, M.~S. Ghaemi, H.~A. Soufiani, and R.~Sundaram, ``Chiron: A robust recommendation system with graph regularizer,'' in \emph{Proceedings of the 10th International Conference on Computer Recognition Systems CORES 2017 10}.\hskip 1em plus 0.5em minus 0.4em\relax Springer, 2018, pp. 367--376.

\bibitem{wu2020robust}
D.~Wu, G.~Lu, and Z.~Xu, ``Robust and accurate representation learning for high-dimensional and sparse matrices in recommender systems,'' in \emph{2020 IEEE International Conference on Knowledge Graph (ICKG)}.\hskip 1em plus 0.5em minus 0.4em\relax IEEE, 2020, pp. 489--496.

\bibitem{Chen2021}
H.~Chen, L.~Wang, Y.~Lin, C.-C.~M. Yeh, F.~Wang, and H.~Yang, ``Structured graph convolutional networks with stochastic masks for recommender systems,'' in \emph{Proceedings of the 44th International ACM SIGIR Conference on Research and Development in Information Retrieval}.\hskip 1em plus 0.5em minus 0.4em\relax Association for Computing Machinery, 2021, pp. 614--623.

\bibitem{Chen2020}
J.~Chen, H.~Xu, J.~Wang, Q.~Xuan, and X.~Zhang, ``Adversarial detection on graph structured data,'' in \emph{Proceedings of the 2020 Workshop on Privacy-Preserving Machine Learning in Practice}.\hskip 1em plus 0.5em minus 0.4em\relax Association for Computing Machinery, 2020, pp. 37--41.

\bibitem{Zhu2019}
D.~Zhu, Z.~Zhang, P.~Cui, and W.~Zhu, ``Robust graph convolutional networks against adversarial attacks,'' in \emph{Proceedings of the 25th ACM SIGKDD International Conference on Knowledge Discovery \& Data Mining}.\hskip 1em plus 0.5em minus 0.4em\relax Association for Computing Machinery, 2019, pp. 1399--1407.

\bibitem{wu2021self}
J.~Wu, X.~Wang, F.~Feng, X.~He, L.~Chen, J.~Lian, and X.~Xie, ``Self-supervised graph learning for recommendation,'' in \emph{Proceedings of the 44th international ACM SIGIR conference on research and development in information retrieval}, 2021, pp. 726--735.

\bibitem{zhu2023knowledge}
X.~Zhu, Y.~Du, Y.~Mao, L.~Chen, Y.~Hu, and Y.~Gao, ``Knowledge-refined denoising network for robust recommendation,'' \emph{arXiv preprint arXiv:2304.14987}, 2023.

\bibitem{fan2023mutual}
Z.~Fan, Z.~Liu, H.~Peng, and P.~S. Yu, ``Mutual wasserstein discrepancy minimization for sequential recommendation,'' \emph{arXiv preprint arXiv:2301.12197}, 2023.

\bibitem{he2018adversarial}
X.~He, Z.~He, X.~Du, and T.-S. Chua, ``Adversarial personalized ranking for recommendation,'' in \emph{The 41st International ACM SIGIR conference on research \& development in information retrieval}, 2018, pp. 355--364.

\bibitem{yuan2019adversarial}
F.~Yuan, L.~Yao, and B.~Benatallah, ``Adversarial collaborative neural network for robust recommendation,'' in \emph{Proceedings of the 42nd International ACM SIGIR Conference on Research and Development in Information Retrieval}, 2019, pp. 1065--1068.

\bibitem{chen2019adversarial}
H.~Chen and J.~Li, ``Adversarial tensor factorization for context-aware recommendation,'' in \emph{Proceedings of the 13th ACM Conference on Recommender Systems}, 2019, pp. 363--367.

\bibitem{wu2021fight}
C.~Wu, D.~Lian, Y.~Ge, Z.~Zhu, E.~Chen, and S.~Yuan, ``Fight fire with fire: towards robust recommender systems via adversarial poisoning training,'' in \emph{Proceedings of the 44th International ACM SIGIR Conference on Research and Development in Information Retrieval}, 2021, pp. 1074--1083.

\bibitem{tang2019adversarial}
J.~Tang, X.~Du, X.~He, F.~Yuan, Q.~Tian, and T.-S. Chua, ``Adversarial training towards robust multimedia recommender system,'' \emph{IEEE Transactions on Knowledge and Data Engineering}, vol.~32, no.~5, pp. 855--867, 2019.

\bibitem{paul2022robust}
A.~Paul, Z.~Wu, K.~Liu, and S.~Gong, ``Robust multi-objective visual bayesian personalized ranking for multimedia recommendation,'' \emph{Applied Intelligence}, pp. 1--12, 2022.

\bibitem{quan2023robust}
Y.~Quan, J.~Ding, C.~Gao, L.~Yi, D.~Jin, and Y.~Li, ``Robust preference-guided denoising for graph based social recommendation,'' in \emph{Proceedings of the ACM Web Conference 2023}, 2023, pp. 1097--1108.

\bibitem{wang2023denoised}
T.~Wang, L.~Xia, and C.~Huang, ``Denoised self-augmented learning for social recommendation,'' \emph{arXiv preprint arXiv:2305.12685}, 2023.

\bibitem{jia2013robust}
D.~Jia, F.~Zhang, and S.~Liu, ``A robust collaborative filtering recommendation algorithm based on multidimensional trust model.'' \emph{J. Softw.}, vol.~8, no.~1, pp. 11--18, 2013.

\bibitem{jia2014robust}
D.~Jia and F.~Zhang, ``A robust collaborative recommendation algorithm incorporating trustworthy neighborhood model.'' \emph{J. Comput.}, vol.~9, no.~10, pp. 2328--2334, 2014.

\bibitem{gao2014robust}
M.~Gao, B.~Ling, Q.~Yuan, Q.~Xiong, L.~Yang \emph{et~al.}, ``A robust collaborative filtering approach based on user relationships for recommendation systems,'' \emph{Mathematical Problems in Engineering}, vol. 2014, 2014.

\bibitem{yi2016robust}
H.~Yi and F.~Zhang, ``Robust recommendation method based on suspicious users measurement and multidimensional trust,'' \emph{Journal of Intelligent Information Systems}, vol.~46, no.~2, pp. 349--367, 2016.

\bibitem{yu2019robust}
H.~Yu, L.~Sun, and F.~Zhang, ``A robust bayesian probabilistic matrix factorization model for collaborative filtering recommender systems based on user anomaly rating behavior detection.'' \emph{KSII Transactions on Internet \& Information Systems}, vol.~13, no.~9, 2019.

\bibitem{zhang2020gcn}
S.~Zhang, H.~Yin, T.~Chen, Q.~V.~N. Hung, Z.~Huang, and L.~Cui, ``Gcn-based user representation learning for unifying robust recommendation and fraudster detection,'' in \emph{Proceedings of the 43rd international ACM SIGIR conference on research and development in information retrieval}, 2020, pp. 689--698.

\bibitem{wang2022towards}
Q.~Wang, D.~Lian, C.~Wu, and E.~Chen, ``Towards robust recommender systems via triple cooperative defense,'' in \emph{International Conference on Web Information Systems Engineering}.\hskip 1em plus 0.5em minus 0.4em\relax Springer, 2022, pp. 564--578.

\bibitem{Yang2021}
Y.~Yang, L.~Wu, R.~Hong, K.~Zhang, and M.~Wang, ``Enhanced graph learning for collaborative filtering via mutual information maximization,'' in \emph{Proceedings of the 44th International ACM SIGIR Conference on Research and Development in Information Retrieval}.\hskip 1em plus 0.5em minus 0.4em\relax Association for Computing Machinery, 2021, pp. 71--80.

\bibitem{Du2023}
Y.~Du, X.~Zhu, L.~Chen, Z.~Fang, and Y.~Gao, ``Metakg: Meta-learning on knowledge graph for cold-start recommendation,'' \emph{IEEE Transactions on Knowledge and Data Engineering}, vol.~35, no.~10, pp. 9850--9863, 2023.

\bibitem{gantner2012personalized}
Z.~Gantner, L.~Drumond, C.~Freudenthaler, and L.~Schmidt-Thieme, ``Personalized ranking for non-uniformly sampled items,'' in \emph{Proceedings of KDD Cup 2011}.\hskip 1em plus 0.5em minus 0.4em\relax PMLR, 2012, pp. 231--247.

\bibitem{zhang2013optimizing}
W.~Zhang, T.~Chen, J.~Wang, and Y.~Yu, ``Optimizing top-n collaborative filtering via dynamic negative item sampling,'' in \emph{Proceedings of the 36th international ACM SIGIR conference on Research and development in information retrieval}, 2013, pp. 785--788.

\bibitem{wang2021implicit}
Z.~Wang, Q.~Xu, Z.~Yang, X.~Cao, and Q.~Huang, ``Implicit feedbacks are not always favorable: Iterative relabeled one-class collaborative filtering against noisy interactions,'' in \emph{Proceedings of the 29th ACM International Conference on Multimedia}, 2021, pp. 3070--3078.

\bibitem{zhu2022gain}
Q.~Zhu, H.~Zhang, Q.~He, and Z.~Dou, ``A gain-tuning dynamic negative sampler for recommendation,'' in \emph{Proceedings of the ACM Web Conference 2022}, 2022, pp. 277--285.

\bibitem{gao2022self}
Y.~Gao, Y.~Du, Y.~Hu, L.~Chen, X.~Zhu, Z.~Fang, and B.~Zheng, ``Self-guided learning to denoise for robust recommendation,'' in \emph{Proceedings of the 45th International ACM SIGIR Conference on Research and Development in Information Retrieval}, 2022, pp. 1412--1422.

\bibitem{chen2019improving}
J.~Chen, D.~Lian, and K.~Zheng, ``Improving one-class collaborative filtering via ranking-based implicit regularizer,'' in \emph{Proceedings of the AAAI Conference on artificial intelligence}, vol.~33, no.~01, 2019, pp. 37--44.

\bibitem{wang2021denoising}
W.~Wang, F.~Feng, X.~He, L.~Nie, and T.-S. Chua, ``Denoising implicit feedback for recommendation,'' in \emph{Proceedings of the 14th ACM international conference on web search and data mining}, 2021, pp. 373--381.

\bibitem{wang2022learning}
Y.~Wang, X.~Xin, Z.~Meng, J.~M. Jose, F.~Feng, and X.~He, ``Learning robust recommenders through cross-model agreement,'' in \emph{Proceedings of the ACM Web Conference 2022}, 2022, pp. 2015--2025.

\bibitem{wang2023efficient}
Z.~Wang, M.~Gao, W.~Li, J.~Yu, L.~Guo, and H.~Yin, ``Efficient bi-level optimization for recommendation denoising,'' in \emph{Proceedings of the 29th ACM SIGKDD Conference on Knowledge Discovery and Data Mining}, 2023, pp. 2502--2511.

\bibitem{yang2018robust}
P.~Yang, P.~Zhao, V.~W. Zheng, L.~Ding, and X.~Gao, ``Robust asymmetric recommendation via min-max optimization,'' in \emph{The 41st International ACM SIGIR Conference on Research \& Development in Information Retrieval}, 2018, pp. 1077--1080.

\bibitem{harper2015movielens}
F.~M. Harper and J.~A. Konstan, ``The movielens datasets: History and context,'' \emph{Acm transactions on interactive intelligent systems (tiis)}, vol.~5, no.~4, pp. 1--19, 2015.

\bibitem{tang2012etrust}
J.~Tang, H.~Gao, H.~Liu, and A.~Das~Sarma, ``etrust: Understanding trust evolution in an online world,'' in \emph{Proceedings of the 18th ACM SIGKDD international conference on Knowledge discovery and data mining}, 2012, pp. 253--261.

\bibitem{10.5555/2540128.2540506}
G.~Guo, J.~Zhang, and N.~Yorke-Smith, ``A novel bayesian similarity measure for recommender systems,'' in \emph{Proceedings of the Twenty-Third International Joint Conference on Artificial Intelligence}, ser. IJCAI '13.\hskip 1em plus 0.5em minus 0.4em\relax AAAI Press, 2013, p. 2619–2625.

\bibitem{guo2014etaf}
G.~Guo, J.~Zhang, D.~Thalmann, and N.~Yorke-Smith, ``Etaf: An extended trust antecedents framework for trust prediction,'' in \emph{Proceedings of the 2014 International Conference on Advances in Social Networks Analysis and Mining (ASONAM)}, 2014, pp. 540--547.

\bibitem{ziegler2005improving}
C.-N. Ziegler, S.~M. McNee, J.~A. Konstan, and G.~Lausen, ``Improving recommendation lists through topic diversification,'' in \emph{Proceedings of the 14th international conference on World Wide Web}, 2005, pp. 22--32.

\bibitem{geng2015learning}
X.~Geng, H.~Zhang, J.~Bian, and T.-S. Chua, ``Learning image and user features for recommendation in social networks,'' in \emph{Proceedings of the IEEE international conference on computer vision}, 2015, pp. 4274--4282.

\bibitem{cho2011friendship}
E.~Cho, S.~A. Myers, and J.~Leskovec, ``Friendship and mobility: user movement in location-based social networks,'' in \emph{Proceedings of the 17th ACM SIGKDD international conference on Knowledge discovery and data mining}, 2011, pp. 1082--1090.

\bibitem{gulla2017adressa}
J.~A. Gulla, L.~Zhang, P.~Liu, {\"O}.~{\"O}zg{\"o}bek, and X.~Su, ``The adressa dataset for news recommendation,'' in \emph{Proceedings of the international conference on web intelligence}, 2017, pp. 1042--1048.

\bibitem{wang2021robust}
F.~Wang, H.~Zhu, G.~Srivastava, S.~Li, M.~R. Khosravi, and L.~Qi, ``Robust collaborative filtering recommendation with user-item-trust records,'' \emph{IEEE Transactions on Computational Social Systems}, vol.~9, no.~4, pp. 986--996, 2021.

\bibitem{li2022causal}
Y.~Li, H.~Chen, J.~Tan, and Y.~Zhang, ``Causal factorization machine for robust recommendation,'' in \emph{Proceedings of the 22nd ACM/IEEE Joint Conference on Digital Libraries}, 2022, pp. 1--9.

\bibitem{fulan2018robust}
Q.~Fulan, Y.~Ruxia, Z.~Shu, and Z.~Yanping, ``Robust recommendation algorithm using an iterative group-based reputation,'' in \emph{2018 IEEE 3rd International Conference on Cloud Computing and Big Data Analysis (ICCCBDA)}.\hskip 1em plus 0.5em minus 0.4em\relax IEEE, 2018, pp. 62--70.

\bibitem{noh2015auro}
G.~Noh and H.~Oh, ``Auro-rec: An unsupervised and robust sybil attack defense in online recommender systems,'' in \emph{2015 SAI Intelligent Systems Conference (IntelliSys)}.\hskip 1em plus 0.5em minus 0.4em\relax IEEE, 2015, pp. 1017--1024.

\bibitem{kaur2016shilling}
P.~Kaur and S.~Goel, ``Shilling attack models in recommender system,'' in \emph{2016 International conference on inventive computation technologies (ICICT)}, vol.~2.\hskip 1em plus 0.5em minus 0.4em\relax IEEE, 2016, pp. 1--5.

\bibitem{wu2022neural}
C.~Wu, R.~Zhang, J.~Guo, Y.~Fan, and X.~Cheng, ``Are neural ranking models robust?'' \emph{ACM Transactions on Information Systems}, vol.~41, no.~2, pp. 1--36, 2022.

\bibitem{wangRecADUnifiedLibrary2023}
{\relax CHANGSHENG}.~WANG, J.~Ye, W.~Wang, C.~Gao, F.~Feng, and X.~He, ``Recad: Towards a unified library for recommender attack and defense,'' in \emph{Proceedings of the 17th ACM Conference on Recommender Systems}.\hskip 1em plus 0.5em minus 0.4em\relax Association for Computing Machinery, 2023, pp. 234--244.

\bibitem{wangPoisoningAttacksRecommender2024}
Z.~Wang, M.~Gao, J.~Yu, H.~Ma, H.~Yin, and S.~Sadiq, ``Poisoning attacks against recommender systems: A survey,'' 2024.

\bibitem{koren2009matrix}
Y.~Koren, R.~Bell, and C.~Volinsky, ``Matrix factorization techniques for recommender systems,'' \emph{Computer}, vol.~42, no.~8, pp. 30--37, 2009.

\bibitem{He2017}
X.~He, L.~Liao, H.~Zhang, L.~Nie, X.~Hu, and T.-S. Chua, ``Neural collaborative filtering,'' in \emph{Proceedings of the 26th International Conference on World Wide Web}.\hskip 1em plus 0.5em minus 0.4em\relax International World Wide Web Conferences Steering Committee, 2017, pp. 173--182.

\bibitem{heLightGCNSimplifyingPowering2020}
X.~He, K.~Deng, X.~Wang, Y.~Li, Y.~Zhang, and M.~Wang, ``Lightgcn: Simplifying and powering graph convolution network for recommendation,'' in \emph{Proceedings of the 43rd International ACM SIGIR Conference on Research and Development in Information Retrieval}.\hskip 1em plus 0.5em minus 0.4em\relax Association for Computing Machinery, 2020, pp. 639--648.

\bibitem{laiAdversariallyRobustRecommendation2023}
Y.~Lai, Y.~Zhu, W.~Fan, X.~Zhang, and K.~Zhou, ``Towards adversarially robust recommendation from adaptive fraudster detection,'' \emph{IEEE Transactions on Information Forensics and Security}, pp. 1--1, 2023.

\end{thebibliography}

\vspace{-11 mm}

\begin{IEEEbiography}[{\includegraphics[width=1in,height=1.25in,clip,keepaspectratio]{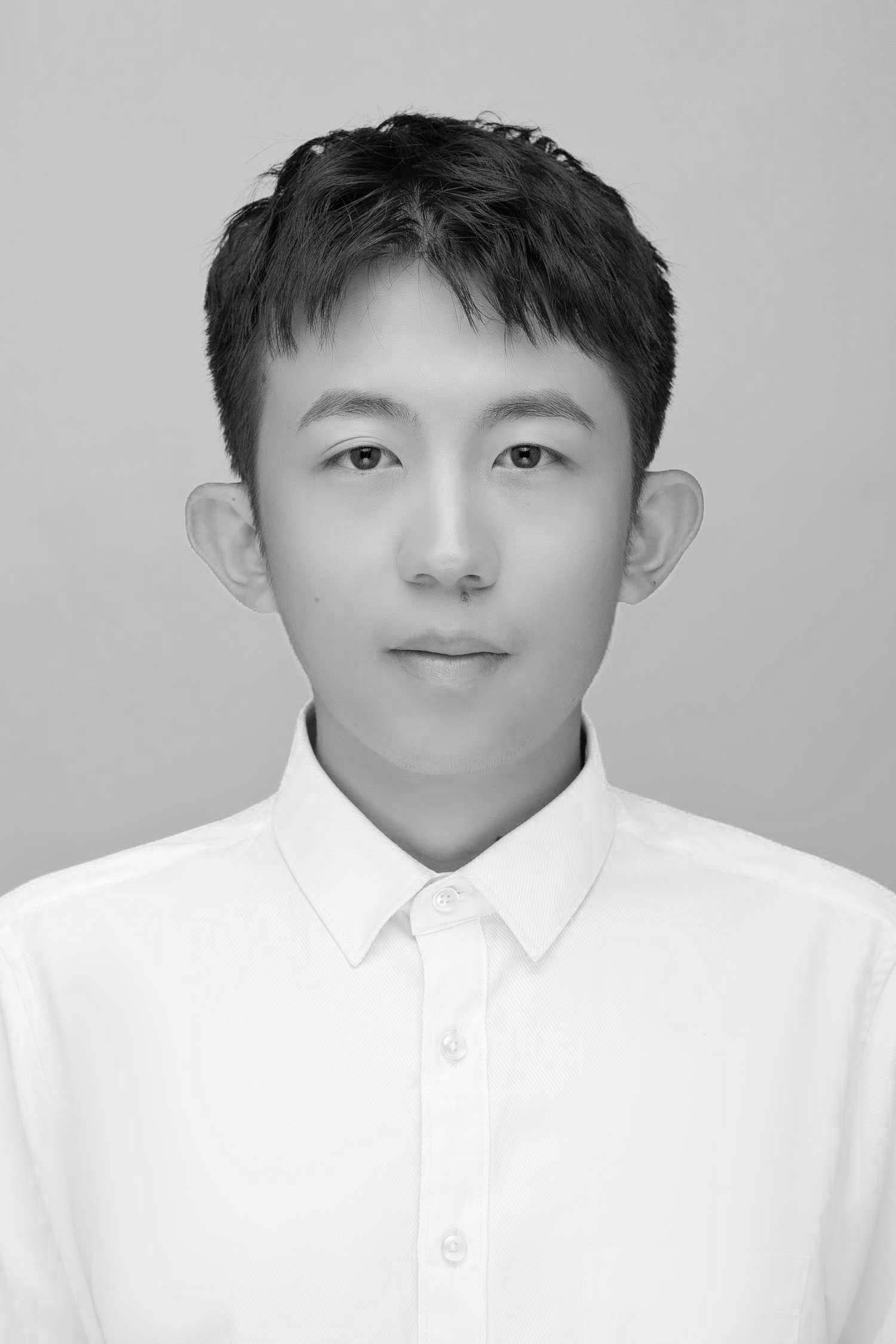}}]{Lei Cheng}
received the BS degree in Data Science and Big Data Technologies from the China University of Mining and Technology, in 2022. He is
currently working toward the master degree in the School of Beijing Jiaotong University. His current research interests
include data mining and recommender systems.
\end{IEEEbiography}
\vspace{-13 mm} 

\begin{IEEEbiography}[{\includegraphics[width=1in,height=1.25in,clip,keepaspectratio]{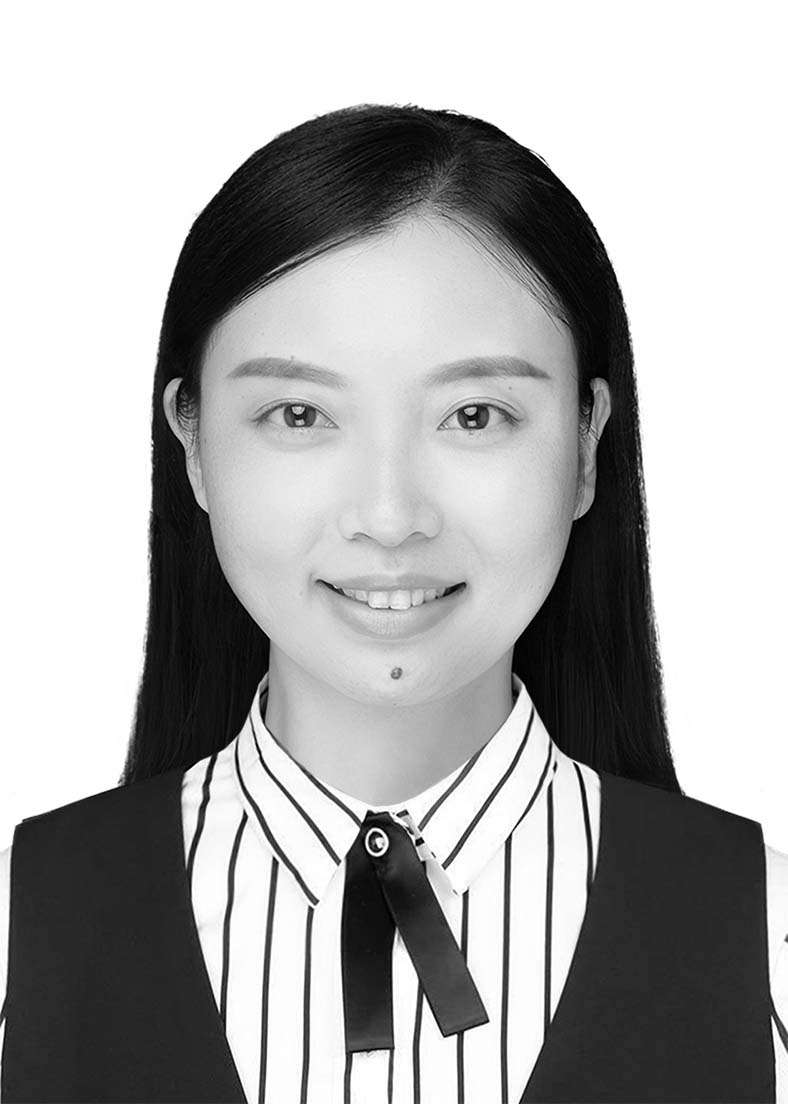}}]{Xiaowen Huang}
 received her BS degree from Central South University in 2015, and her Ph.D. degree from Institute of Automation, Chinese Academy of Sciences (CASIA) in 2020. She is currently an associate professor at Beijing Jiaotong University. Her research interest is in multimedia analysis and data mining. Her research works are published in the top international conferences and journals, including ACM Multimedia, CVPR, WWW, TPAMI, TKDE, TOIS, TOMM, and so on.
\end{IEEEbiography}
\vspace{-13 mm} 

\begin{IEEEbiography}[{\includegraphics[width=1in,height=1.25in,clip,keepaspectratio]{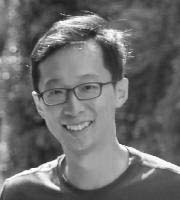}}]{Jitao Sang}
received the BS degre in computer science and technology from Southeast university, China and the PhD degree in Pattern Recognition from University of Chinese Academy of Sciences, National Lab of Pattern Recognition, Institute of Automation,China. He is currently a professor with Beijing Jiaotong University. His current research intersts include Machine Learning and Cognitive Computing, Artificial Intelligence and Applications and Computer Technolog.
\end{IEEEbiography}
\vspace{-13 mm} 
\begin{IEEEbiography}[{\includegraphics[width=1in,height=1.25in,clip,keepaspectratio]{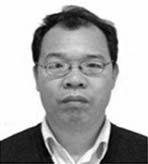}}]{Jian Yu}
received the BS and PhD degrees in mathematical sciences from Peking University, China. He is currently a professor with the Beijing Jiaotong University, He is a director of China artificial intelligence society, a standing member of machine learning professional committee of China artificial intelligence society and a member of IEEE. His current research interests include Machine Learning and Cognitive Computing, Artificial Intelligence and Applications and Network Information Security.
\end{IEEEbiography}
\vspace{-13 mm} 

\vfill

\end{document}